\documentclass[journal]{IEEEtran}

\usepackage{cite}
\ifCLASSINFOpdf
\fi
\usepackage{algorithm}
\usepackage{algpseudocode}
\usepackage{amsmath,amsfonts,amsthm} 
\usepackage{amssymb}
\usepackage{tikz}
\usepackage{pgfplots}
\usepgfplotslibrary{fillbetween}
\usepackage{float}
\usepackage{bbm}
\usepackage{float}
\usepackage{graphicx}
\usepackage{algorithm}
\usepackage{algpseudocode}
\usepackage{subcaption}
\usepackage{siunitx}
\usepackage{tikz}
\usetikzlibrary{arrows,automata}
\usepackage{multicol}
\usepackage{comment}
\usepackage{pgfplots}
\usepackage{epstopdf}
\usepackage{multirow}
\usepackage[font=scriptsize,skip=5pt]{caption}
\usepackage{soul}
\usepackage{todonotes}
\usepackage{color, colortbl}
\usepackage{array}
\usetikzlibrary{patterns}
\usepackage{balance}

\usepackage{enumitem}
\usepackage{footmisc}
\usepackage{pgfplots}
\usepackage{caption}
\usepackage{subcaption}
\usepackage{wrapfig}
\newtheorem{theorem}{Theorem}

\newtheorem{lemma}[theorem]{Lemma}

\allowdisplaybreaks
\begin{document}
	\setlength{\parskip}{0em}
	
	\title{Version Age of Information Minimization over Fading Broadcast Channels}
\author{\IEEEauthorblockN{Gangadhar Karevvanavar\IEEEauthorrefmark{1}, Hrishikesh Pable\IEEEauthorrefmark{1}, Om Patil\IEEEauthorrefmark{1},  Rajshekhar V Bhat\IEEEauthorrefmark{1} and Nikolaos Pappas\IEEEauthorrefmark{2}}
	
	\IEEEauthorblockA{\IEEEauthorrefmark{1}Indian Institute of Technology Dharwad, Dharwad, Karnataka, India, \\
 \IEEEauthorrefmark{2}Department of Computer and Information Science Link\"oping University, Sweden\\
		Email: 
  \IEEEauthorrefmark{1}\{212021007, 200010036, 200010037, rajshekhar.bhat\}@iitdh.ac.in,  
		\IEEEauthorrefmark{2}nikolaos.pappas@liu.se
}}
	\maketitle

 \pagenumbering{gobble} 
 
\begin{abstract}
We consider a base station (BS) that receives version update packets from multiple exogenous streams and broadcasts them to corresponding users over a fading broadcast channel using a non-orthogonal multiple access (NOMA) scheme. Sequentially indexed packets arrive randomly in each stream, with new packets making the previous ones obsolete. In this case, we consider the version age of information (VAoI) at a user, defined as the difference in the version index of the latest available packet at the BS and that at the user, as a metric of freshness of information. Our objective is to minimize a weighted sum of average VAoI across users subject to an average power constraint at the BS by optimally scheduling the update packets from various streams for transmission and transmitting them with sufficient powers to guarantee their successful delivery. We consider the class of channel-only stationary randomized policies (CO-SRP), which rely solely on channel power gains for transmission decisions. We solve the resulting non-convex problem optimally and show that the VAoI achieved under the optimal CO-SRP is within twice the optimal achievable VAoI. We also obtained a Constrained Markov Decision Process (CMDP)-based solution and its structural properties. Numerical simulations show a close performance between the optimal CO-SRP and CMDP-based solutions. Additionally, a time division multiple access (TDMA) scheme, which allows transmission to at most one user at a time, matches NOMA's performance under tight average power constraints. However, NOMA outperforms TDMA as the constraint is relaxed.
\end{abstract}
	
\IEEEpeerreviewmaketitle

\section{Introduction}
In the emerging communication networks, fresh information delivery is of great importance. As such, researchers have made tremendous efforts over the past decade to quantify, analyze, and optimize the freshness of information over networks. Metrics such as the age of information (AoI) and its variants have been considered and optimized in various communication scenarios \cite{Book_Nikos, Book_Yates, pappas2023age}.  

Recently, a metric called Version Age of Information (VAoI), defined as the difference in the version indices of the latest information packets available at the transmitter and receiver, has been considered \cite{VAoI_Yates, VAoI_Ulukus, delfani2023version}. Unlike the AoI, which starts to increase after receiving a packet irrespective of whether or not there is a new packet available at the source, the VAoI increases only when a packet arrives at the transmitter. We consider the optimization of the VAoI metric in a downlink broadcast channel (BC), in which a single base station (BS) transmits information to multiple receivers simultaneously, using non-orthogonal multiple access (NOMA). 
 
In the literature, optimization of AoI and its variants over BCs have been considered under various assumptions \cite{BCN_Book_Chapter, MU_OMA_NOMA, Hybrid_OMA_NOMA, Non_stationary, Coding_AoI, Downlink_NOMA, BWN_Modiano, BWN_Modiano_2, Age_of_Synchronization}.  
The works \cite{MU_OMA_NOMA, Hybrid_OMA_NOMA, Non_stationary, Coding_AoI, Downlink_NOMA}  and \cite{BWN_Modiano, BWN_Modiano_2, Age_of_Synchronization} consider optimization of AoI over BCs under NOMA and time-division multiple access (TDMA) schemes, respectively. 
Communications over noiseless BC \cite{Modiano_No_Buffer_Case, Noiseless_BC, With_CSI, BroadCast_MIMO} and erasure BC \cite{BC-Erasure} have been considered. The works   \cite{BWN_Modiano, BWN_Modiano_2, Index_Policy,kadota2019_1, Coding_AoI, BC-Erasure} consider simple on-off fading channel where a transmitted packet is considered to be successful with some probability and fails otherwise, and  \cite{MU_OMA_NOMA, Hybrid_OMA_NOMA, Non_stationary, With_CSI} consider a more general fading channel where the channel power gain realizations can take more than two states and the probability of success of a transmission depends on the transmission power.  
Different levels of knowledge about channel power gains at the BS have been assumed. For instance,  \cite{MU_OMA_NOMA, Hybrid_OMA_NOMA} consider a broadcast channel in which the BS has statistical knowledge of channel state information only, without perfect channel state information at the transmitter (CSIT), and \cite{Downlink_NOMA, Non_stationary, With_CSI, BroadCast_MIMO} consider AoI minimization with perfect CSIT at the BS. 
In the above works, AoI minimization is studied using different tools. For instance,  \cite{MU_OMA_NOMA,Hybrid_OMA_NOMA,Index_Policy,Modiano_No_Buffer_Case,Age_of_Synchronization} adopt the Markov Decision Process (MDP) framework,  \cite{Downlink_NOMA} adopts a Lyapunov optimization based drift-plus-penalty framework. Index-based  \cite{BWN_Modiano,BWN_Modiano_2,Index_Policy,Modiano_No_Buffer_Case,Noiseless_BC,Age_of_Synchronization} and stationary randomized policies \cite{BWN_Modiano,BWN_Modiano_2,kadota2019_1,Coding_AoI} have also been considered. The works, such as \cite{OMA_NOMA} and \cite{AoB}, analyze the average AoI. 

In some of the above works, including   \cite{MU_OMA_NOMA, Hybrid_OMA_NOMA}, both user scheduling and transmit power control are considered, as in the current work. However, our work differs from these, as we consider perfect CSIT, as in \cite{Non_stationary, Downlink_NOMA, With_CSI, BroadCast_MIMO} and adopt the NOMA scheme. In applying NOMA with perfect CSIT, we employ superposition coding at the transmitter and successive interference cancellation at the receiver, with the decoding order determined by the channel power gains of all users. 
In this case, the transmit powers that are required to reliably deliver a required number of bits of packets depend on the channel power gain realizations and decoding order. As the channel realizations can change over slots, the decoding order also needs to be changed in every slot. Thus, the power required for the successful delivery of update packets varies in every slot, in turn impacting the streams that should be scheduled for transmission in a slot. In \cite{Non_stationary}, non-stationary channels have been considered, and in \cite{Downlink_NOMA}, only two users are considered, in which the channel power gain of a user is assumed to be always greater than the other, due to which the set of users and their decoding order need not be changed in every slot. The works \cite{With_CSI, BroadCast_MIMO}, which consider perfect CSIT, obtain optimal precoding schemes for AoI minimization. 
Moreover, almost all the above works, including \cite{MU_OMA_NOMA, Hybrid_OMA_NOMA, Downlink_NOMA, BroadCast_MIMO,With_CSI, OMA_NOMA,AoB} consider the AoI as the metric,  unlike in the current work where VAoI is considered as the metric. 

In summary, none of the earlier works consider the minimization of VAoI over a BC with perfect CSIT at the BS using NOMA, which we consider in the current work. The results from earlier works, even those that consider NOMA with perfect CSIT, cannot be generalized or specialized to obtain results for the considered problem. Our main contributions are as follows: 
\begin{itemize}[leftmargin=*]
\item To minimize a long-term expected average weighted sum of  VAoI across users subject to an average transmit power constraint, we consider a class of channel-only stationary randomized policies (CO-SRP), in which the transmission decisions are made based only on the channel power gain realizations. We construct the induced Markov chain and derive its stationary distribution to obtain the optimal CO-SRP. Using this, we express the average VAoI and average power under the CO-SRP and reformulate the original problem. The reformulated problem involves a sum of affine ratios, making it non-convex. 
\item Based on approaches adopted in prior works, we propose an algorithm to solve the reformulated problem and obtain the optimal CO-SRP. We show that the VAoI achieved by the optimal CO-SRP is within twice the optimal VAoI. 
\item By considering the VAoI and channel power gains of all the users as a state, we obtain a CMDP-based solution by Lagrangian relaxation. We also obtain the structural properties of the solution to the Lagrangian relaxed problem. 
\item Through numerical simulations, we show that our proposed CO-SRP solution performs competitively with the CMDP-based approach while being significantly simpler. Furthermore, a TDMA scheme allows transmission to only one user at a time and achieves comparable performance to NOMA when the bound on the average power constraint is low. However, as the bound is increased, NOMA outperforms TDMA.
\end{itemize}

\section{System Model}\label{sec:sys_model}
We consider a BS that receives status update packets from $N$ streams to be communicated to $N$ destination users. The BS  consists of a transmitter and $N$ \emph{single-packet} queues, each for storing the latest packet of a user, where a queue can store at most one packet. Time is slotted with slot index $t \in \{1, 2, ...\}$. Below, we describe the packet arrival model, channel model, and the NOMA strategy considered, followed by a description of the performance metric adopted and the problem formulation. 

\subsection{Packet Arrival Model}
At the beginning of every slot $t$, a new packet from stream $i \in \mathcal{N}\triangleq \{1, 2, ..., N\}$ arrives with a certain probability, the packet needs to be delivered to User $i \in \mathcal{N}$. Let $A_i(t) \in \{0, 1\}$ be the indicator function that is equal to $1$ when a packet from the $i^{\rm th}$ stream arrives in slot $t$, and $A_i(t) = 0$ otherwise. The packet arrival follows a Bernoulli process that is independent and identically distributed (i.i.d.) across time, with $\mathbb{P}(A_i(t) = 1) = \lambda_i,\; \forall i\in \mathcal{N}$ and $t\in \{1,2,\ldots\}$. 
The latest packet from the $i^{\rm th}$ stream is stored in Queue $i$. We consider that each packet arriving at the $i^{\rm th}$ stream contains $R^0_i$ bits. 

\subsection{Channel Model and NOMA Strategy}
Let $H_i(t)\in \mathcal{H}$  be the channel power gain between the BS and User $i$ in slot $t$ for all $i \in \mathcal{N}$.  
We consider that $H_i(1),H_i(2)\ldots,$ are i.i.d. for each User $i$ and that $\mathcal{H}$ is a finite set with positive  elements.  
Let $u_i(t) \in \{0, 1\}$ be the indicator function, which equals $1$ when the BS transmits a packet from the $i^{\rm th}$ stream during slot $t$, and $u_i(t) = 0$ otherwise. For transmission, the BS adopts superposition coding, which allocates power $P_i(t)$ to the message corresponding to User $i$ in time slot $t$, encodes and superimposes the codewords, and transmits them over the fading BC. Each user employs successive interference cancellation (SIC) to decode its message from the received symbols. Under this strategy, we describe the relationship between transmit powers and achievable rates next.

Let $O_1(t)$ be the index of the user with the highest channel power gain, $O_2(t)$  be the index of the user with the next highest channel power gain, and so on, in slot $t$. That is, $h_{O_1}(t)\geq h_{O_2}(t)\geq \ldots h_{O_N}(t)$ is satisfied.\footnote{Multiple orderings are possible if channel power gain realizations of more than one user are same. In such cases, we consider the ordering that places the user with a lower index before the user with a higher index having the same channel power gain.} We denote $O^{-1}_i(t)$ as the position of User $i$ when the users are arranged according to decreasing values of their channel power gains.
As mentioned, the users adopt SIC for decoding, where the user with index $O_N(t)$ decodes its message by considering codewords from users $O_{N-1}(t),\ldots, O_1(t)$ as noise, the User $O_{N-1}(t)$ first decodes the message of User $O_N(t)$ (which is possible because the channel for User $O_{N-1}(t)$ is better than that for User $O_N(t)$), subtracts the encoded version of the message from the received symbols and then decodes its message by considering the codewords from users $O_{N-2}(t),\ldots, O_1(t)$ as noise. The users $O_{N-2}(t), O_{N-3}(t),\ldots, 2, 1$, decode the messages in the similar manner \cite{Cover_Thomas,Sumei_Sum_BC_Region}.  

Under the superposition coding at the BS and SIC at the users, the signal-to-interference-plus-noise ratio, considering unit noise variance, experienced at User $i$ is given by 
\begin{align*}
    S_i(t) = \frac{P_i(t)h_i(t)}{{1+ h_i(t)\sum_{k<O^{-1}_i(t)}P_{O_k}(t)}},
\end{align*}
where $h_i(t)$ is the realization of the random channel power gain $H_i(t)$. In this case, we consider that the maximum achievable rate for User $i$ is given by $f(S_i(t))$, where $f(\cdot)$ is a concave, non-decreasing function with $f(0)=0$ \cite{Sumei_Sum_BC_Region}. 
In the above, when transmitting a packet for User $i$, it is optimal to transmit a minimum of $R^0_i$ bits of the packet; otherwise, the update will not be received successfully. Hence, the power consumed for transmitting less than $R^0_i$ bits does not result in any reward, or in other words, the reward earned will be the same as when no bits are transmitted.  
Hence, with the knowledge of the channel power gains of all users, the BS allocates powers such that it delivers all the bits of a packet. Thus, $P_{O_1}, P_{O_2},\ldots,$ are chosen such that
\begin{align*}
f(h_{O_1}P_{O_1}u_{O_1}) &= R^0_{O_1}u_{O_1},\\
f\left(\frac{h_{O_2}P_{O_2}u_{O_2}}{1+h_{O_2}P_{O_1}u_{O_1}}\right) &= R^0_{O_2}u_{O_2},\\
&\ldots\nonumber \\
f\left(\frac{h_{O_i}P_{O_i}u_{O_i}}{1+h_{O_i}\sum_{k=1}^{i-1} P_{O_k}u_{O_k}}\right) &= R^0_{O_i}u_{O_i},
\end{align*}
and so on until $i=N$.  
From the above, we obtain
\begin{flalign}\label{eq:power}
&P_{O_i}u_{O_i} = \frac{f^{-1}\left( R^{0}_{O_i} u_{O_i} \right)}{h_{O_i}} + f^{-1}\left( R^{0}_{O_i} u_{O_i} \right) \sum_{k=1}^{i-1} P_{O_k}u_{O_k},&&\nonumber\\
&\text{for all }  i=1,2,\ldots,N. 
\end{flalign}

\subsection{Performance Metric and Problem Formulation}
Let $z_i (t)$ denote the version of the packet in the $i^{\rm th}$ queue at the beginning of slot $t$. Recall that $A_i(t)$ is the indicator variable that indicates the arrival of a new packet from the $i^{\rm th}$ stream in slot $t$. Then, we have  
\begin{align*}
z_i(t)= \sum_{\tau=1}^{t}A_i(\tau). 
\end{align*}
$z_i(t)$ value changes only when a new packet arrives in the queue. Let $y_i(t)$ be the version of the most recently received packet at Destination $i$ at time $t$. Then, the instantaneous VAoI is defined as $\Delta_i(t) = z_i(t)-y_i(t)$. We measure instantaneous VAoI at the end of a slot.

In this work, we are interested in obtaining transmission scheduling policy, $\pi$, which gives the rule for obtaining the sequence of decisions $\{u_i(t)\}_{i=1}^{N}$ at the BS, for minimization of the long-term expected average VAoI subject an
average power constraint. For given weights $w_1,\ldots,w_N$, this can be accomplished by solving the following optimization problem
\begin{subequations}\label{eq:main-opt-problem}
\begin{align}
V_{\rm opt} = \min_{\pi} \;\lim_{T\rightarrow \infty}&\;\;\frac{1}{T}\sum_{t=1}^{T}\sum_{i=1}^{N}w_i\mathbb{E}[\Delta_i(t)],\\
\text{subject to}&\;\;  \lim_{T\rightarrow \infty}\frac{1}{T}\sum_{t=1}^{T} \sum_{i=1}^N\mathbb{E}[P_i(t)]\leq \bar{P}, \;  \label{eq:power_constraint}
\end{align}
\end{subequations}
Note that for any choice of $\{u_i(t)\}_{i=1}^{N}$, we can obtain the transmit powers required for ensuring all the bits of a packet are delivered from  \eqref{eq:power} for any $t$. 

In this work, we also consider a TDMA scheme, where, at most, one user can transmit in a given time slot. Specifically, we derive the TDMA scheme-based solution by solving \eqref{eq:main-opt-problem} with the additional constraint:
\begin{align}
\sum_{i=1}^N u_i(t)\leq 1, \forall t\in \{1,2,\ldots\}.
\end{align}

Next, we obtain solutions to \eqref{eq:main-opt-problem}. 

\section{Optimization Analysis}\label{sec:solution}
In this section, we first obtain the optimal policy among the class of channel-only stationary randomized policies (CO-SRP) to solve \eqref{eq:main-opt-problem}. We also solve \eqref{eq:main-opt-problem} using the CMDP framework.

 \begin{figure*}[t]
	\centering
\begin{tikzpicture}[->, >=stealth', auto, semithick, node distance=4cm]
\tikzstyle{every state}=[fill=white,draw=black,thick,text=black,scale=1]
\node[state]    (A) 			  {$0$};
\node[state]    (B)[right of=A]   {$1$};
\node[state]    (C)[right of=B]   {$2$};
 \node[state]    (D)[right of=C]   {$n$};
 \node at (10,0) {$\ldots$};
  \node at (14,0) {$\ldots$};

\path

(A) edge[loop left]			node[rotate =90, above, midway,text width=3cm,align=center]{\scriptsize $A_i=0|(A_i =1 \& u_i=1)$\\$1 - \lambda_i (1-  p^{\boldsymbol{\mu}}_i)$}	(A)
(A) edge[bend left,above]	node[text width=3cm,align=center, xshift = -0.3cm]{\scriptsize$A_i=1,u_i=0$\\$\lambda_i (1- p^{\boldsymbol{\mu}}_i)$}	(B)
(B) edge[loop]		node[above, midway,text width=3cm,align=center]{\scriptsize$A_i=0,u_i=0$\\$(1-\lambda_i)(1- p^{\boldsymbol{\mu}}_i)$}	(B)
(B) edge[bend left = 30,below]	node[text width=3cm,align=center, above]{\scriptsize$u_i=1$\\$ p^{\boldsymbol{\mu}}_i$}	(A)
(B) edge[bend left,above]	node[text width=3cm,align=center]{\scriptsize$A_i=1,u_i=0$\\$\lambda_i (1 - p^{\boldsymbol{\mu}}_i)$}	(C)
(C) edge[bend left = 32,below]	node[text width=3cm,align=center, yshift = 0.9cm, xshift=0.7cm]{\scriptsize$u_i=1$\\$p^{\boldsymbol{\mu}}_i$}	(A)
(C) edge[loop]		node[above, midway,text width=3cm,align=center]{\scriptsize$A_i=0,u_i=0$\\$(1-\lambda_i)(1- p^{\boldsymbol{\mu}}_i)$}	(C)
(D) edge[loop]		node[above, midway,text width=3cm,align=center]{\scriptsize$A_i=0,u_i=0$\\$(1-\lambda_i)(1- p^{\boldsymbol{\mu}}_i)$}	(D)
(D) edge[bend left = 33,below]	node[text width=3cm,align=center, yshift = 0.9cm, xshift=0.7cm]{\scriptsize$u_i=1$\\$ p^{\boldsymbol{\mu}}_i$}	(A);

\draw[->,out=45,in=180,dashed] (8.4,0.15) to  (10,0.75);
\draw[->,out=45,in=180,dashed] (12.4,0.15) to  (14,0.75);

\end{tikzpicture}	
\caption{VAoI evolution for the CO-SRP. The expressions closer to the arrows represent the state transition probability; the other expressions above are the events that cause the transitions.}
\label{fig:MC}
\end{figure*}
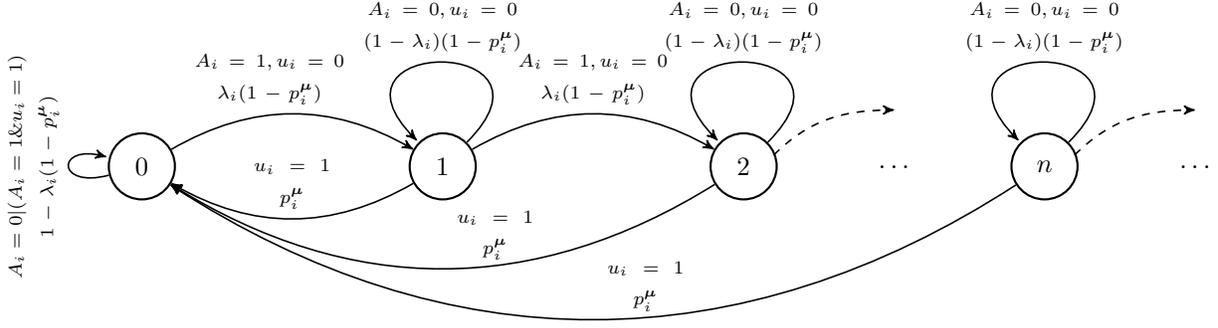 
 
\subsection{Channel-Only Stationary Randomized Policy (CO-SRP)}
Let $2^{\mathcal{N}}$ be the power set of $\mathcal{N}$ and $\mathbf{h} \triangleq (h_1, h_2,\ldots, h_N)$. Our policy is:

\emph{CO-SRP:}
\emph{In a slot, when the channel power gain is $\mathbf{h}$, the BS transmits to users in $\mathcal{W} \in 2^{\mathcal{N}}$ with probability  $\mu_{\mathbf{h}}^{\mathcal{W}}$.}

The policy does not require the BS to have knowledge of the instantaneous VAoI for making transmission decisions, so its execution becomes simple.
Under the policy, the conditional probability of successful delivery of a packet to User $i$, conditioned on $\mathbf{h}$, is $\sum_{\mathcal{W}\in 2^\mathcal{N}:\mathcal{W}\cap \{i\}\neq \phi}\mu_{\mathbf{h}}^{{\mathcal{W}}}$, where $\phi$ is the null set. Hence, the probability of successfully delivering packets to User $i$, 
$p^{\boldsymbol{\mu}}_{i} = \sum_{\mathcal{W}\in 2^\mathcal{N}:\mathcal{W}\cap \{i\}\neq \phi}\mathbb{E}_{\mathbf{H}}[\mu_{\mathbf{h}}^{{\mathcal{W}}}] =   \sum_{\mathbf{h}\in \mathcal{H}^N} \sum_{\mathcal{W}\in 2^\mathcal{N}:\mathcal{W}\cap \{i\}\neq \phi}\mu_{\mathbf{h}}^{{\mathcal{W}}}\;\mathbb{P}(\mathbf{h})$, where $\mathbb{P}(\mathbf{h})$
is the probability of channel being $\mathbf{h}$. 
When the information is transmitted to users, the transmit power allocated for codeword corresponding to User $i \in \mathcal{W}$ when the channel is $\mathbf{h}$, which we denote $P_{i,\mathbf{h}}^{\mathcal{W}}$ can be computed via \eqref{eq:power} and the expected power consumed will be equal to $P^{\boldsymbol{\mu}}_{i} = \sum_{\mathcal{W}\in 2^\mathcal{N}:\mathcal{W}\cap \{i\}\neq \phi}   \mathbb{E}_{\mathbf{H}}[\mu_{\mathbf{h}}^{{\mathcal{W}}} P_{i,\mathbf{h}}^{\mathcal{W}} ]$.
\subsubsection{Algorithm for Obtaining Optimal CO-SRP}
Under this policy, the VAoI evolves as a Markov chain shown in Fig.~\ref{fig:MC}. Below, we derive the long-term expected average VAoI (objective function of \eqref{eq:main-opt-problem}) under the policy by obtaining the stationary distribution of the Markov chain, reformulate \eqref{eq:main-opt-problem} under the CO-SRP, which turns out to be a non-convex sum of affine ratios problem and propose an algorithm to solve it, and obtain a bound on its performance. 
\begin{theorem}\label{thm:SRP_1}
    The long-term expected average VAoI and power under the CO-SRP are given by 
    \begin{align*}
  \lim_{T\rightarrow \infty}\frac{1}{T}\sum_{t=1}^{T}\mathbb{E}[\Delta_i(t)]  = \frac{\lambda_i(1 - p^{\boldsymbol{\mu}}_i)}{p^{\boldsymbol{\mu}}_i},
    \end{align*}
and 
\begin{align*}
    \lim_{T\rightarrow \infty}\frac{1}{T}\sum_{t=1}^{T} \sum_{i=1}^N\mathbb{E}[P_i(t)] = \sum_{i=1}^N \frac{\lambda_i P^{\boldsymbol{\mu}}_i }{\lambda_i ( 1 - p^{\boldsymbol{\mu}}_i) + p^{\boldsymbol{\mu}}_i } ,
\end{align*}
respectively. 
\end{theorem}
\begin{proof}
    Define $\pi_{n,i} \triangleq \mathbb{P}(\Delta_i = n)$ for all $i\in \mathcal{N}$, $n \in \{0, 1, \ldots \}$. Then, we  have
    \begin{align*} 
        \pi_{0,i} &=
            \pi_{0,i} \; (1 - \lambda_i(1 - p^{\boldsymbol{\mu}}_i)) + p^{\boldsymbol{\mu}}_i \sum_{m=1}^{\infty} \pi_{m,i}\\
        \pi_{n,i} &=
            \pi_{n-1,i} \; \lambda_i(1 - p^{\boldsymbol{\mu}}_i) + \pi_{n,i} \; (1 - \lambda_i) (1 - p^{\boldsymbol{\mu}}_i),
    \end{align*}
        for $n \in \{1, 2,\ldots\}$. 
Since $\sum_{n=0}^{\infty} \pi_{n,i} = 1$, we get, 
    \begin{align*}
        &\pi_{n,i} = \frac{p^{\boldsymbol{\mu}}_i}{\lambda_i(1 - p^{\boldsymbol{\mu}}_i) + p^{\boldsymbol{\mu}}_i} \left(\frac{\lambda_i(1 - p^{\boldsymbol{\mu}}_i)}{\lambda_i(1 - p^{\boldsymbol{\mu}}_i) + p^{\boldsymbol{\mu}}_i}\right)^n,
    \end{align*}
    for $n\in \{0, 1, \ldots\}$. 
 Thus, the expected VAoI is 
    \begin{align}\label{eq:expected_VAoI_COSRP}
       \sum_{n=0}^{\infty} n \mathbb{P}(\Delta_i = n)= \sum_{n=0}^{\infty} n \pi_{n,i}
        = \frac{\lambda_i(1 - p^{\boldsymbol{\mu}}_i)}{p^{\boldsymbol{\mu}}_i}.     
    \end{align} 
In computing the expected VAoI, we used the fact that the VAoI transitions from State $0$ to State $0$ under the following events: when no packet arrival occurs ($A_i = 0$), regardless of transmission ($u_i = 1$ or $u_i = 0$), and when an arrival occurs ($A_i = 1$) and the packet is transmitted in the same slot ($u_i = 1$), as shown in Fig.~\ref{fig:MC}. The stationary randomized policy does not consider whether there is a packet in the queue, and the policy can attempt to transmit even when the VAoI is $0$ and no packet arrives. However, transmission is not possible when the queue is empty, i.e., when there is no packet in the queue. To address this, we can modify the stationary randomized policy to depend on the queue state and allow transmission with probability zero when the queue is empty. Another equivalent approach is to permit the transmission even when the queue is empty but account for its non-occurrence by consuming zero power when the policy requests transmission. We adopt the former approach. This is done by subtracting a fraction of $P^{\boldsymbol{\mu}}_i$ equal to the probability of the event that VAoI is in State $0$ times the probability of that no packet arrives, which is equal to $\pi_{0,i} (1-\lambda_i)$. 
By substituting $\pi_{0,i} = p^{\boldsymbol{\mu}}_i/(\lambda_i (1 - p^{\boldsymbol{\mu}}_i) + p^{\boldsymbol{\mu}}_i )$, we obtain 
the probability of queue being non-empty for User $i$ as $(1 -  \pi_{0,i} (1-\lambda_i)) = \lambda_i/(\lambda_i (1 - p^{\boldsymbol{\mu}}_i) + p^{\boldsymbol{\mu}}_i)$. 
Hence, the actual expected average transmit power consumed is given by,
\begin{align}\label{eq:actual_tot_pow}
         &P^{\boldsymbol{\mu}} =\sum_{i=1}^N P^{ \boldsymbol{\mu}}_i \left(1 -  \pi_{0,i} (1-\lambda_i)\right) = \sum_{i=1}^N \frac{\lambda_i P^{ \boldsymbol{\mu}}_i }{\lambda_i ( 1 - p^{\boldsymbol{\mu}}_i) + p^{\boldsymbol{\mu}}_i}. 
\end{align}
\end{proof}
Using Theorem~\ref{thm:SRP_1}, and the definition of $p^{\boldsymbol{\mu}}_{i}$, $P^{ \boldsymbol{\mu}}_i$ in \eqref{eq:expected_VAoI_COSRP} and \eqref{eq:actual_tot_pow}, the optimization problem in \eqref{eq:main-opt-problem} can be reformulated as
\begin{subequations}\label{eq:SRP_new}
\begin{align}
V_{\text{SRP}} = \min_{0 \leq \mu_{\mathbf{h}}^{\mathcal{W}} \leq 1} &\sum_{i=1}^N w_i  O_i(\boldsymbol{\mu}),\label{eq:Obj_fun_11a}\\
\text{subject to}\;\; 
&\sum_{i=1}^N\frac{d_i(\boldsymbol{\mu})}{f_i(\boldsymbol{\mu})} \leq \bar{P},\label{eq:SRP_PowerConstraint}\\
&g_{\mathbf{h}}(\boldsymbol{\mu}) = 0, \quad \forall \mathbf{h} \in \mathcal{H}^N\label{eq:C2_11c},
\end{align}
\end{subequations}
where
\begin{align*}
    \boldsymbol{\mu}& = \{\mu_{\mathbf{h}}^{\mathcal{W}}|\; \forall \mathcal{W} \in 2^{\mathcal{N}}, \forall \mathbf{h} \in \mathcal{H}^N \},\\
O_i(\boldsymbol{\mu}) &= \lambda_i \left( \left(\sum_{\mathbf{h} \in \mathcal{H}^N } \sum_{\mathcal{W} \in 2^\mathcal{N} : \mathcal{W} \cap \{i\} \neq \phi} \mu_{\mathbf{h}}^{\mathcal{W}} \mathbb{P}(\mathbf{h}) \right)^{-1} - 1 \right), \\
d_i(\boldsymbol{\mu}) &= \lambda_i \sum_{\mathbf{h} \in \mathcal{H}^N} \sum_{\mathcal{W} \in 2^\mathcal{N} : \mathcal{W} \cap \{i\} \neq \phi} \mu_{\mathbf{h}}^{\mathcal{W}} P_{i,\mathbf{h}}^{\mathcal{W}}\; \mathbb{P}(\mathbf{h}),\\   
    f_i(\boldsymbol{\mu})& = \lambda_i \left(1 - \sum_{\mathbf{h}\in \mathcal{H}^N} \sum_{\mathcal{W} \in 2^\mathcal{N} : \mathcal{W} \cap \{i\} \neq \phi} \mu_{\mathbf{h}}^{\mathcal{W}} \mathbb{P}(\mathbf{h}) \right) \\
    &\;\;\;+\sum_{\mathbf{h}\in \mathcal{H}^N} \sum_{\mathcal{W} \in 2^\mathcal{N} : \mathcal{W} \cap \{i\} \neq \phi} \mu_{\mathbf{h}}^{\mathcal{W}} \mathbb{P}(\mathbf{h}),\\ 
    g_{\mathbf{h}}(\boldsymbol{\mu}) 
&=  \sum_{\mathcal{W} \in 2^{\mathcal{N}}} \mu_{\mathbf{h}}^{\mathcal{W}} - 1,\; \forall \mathbf{h} \in \mathcal{H}^N,  
\end{align*}
for all $i \in \mathcal{N}$. 
It can be noted that $d_i(\boldsymbol{\mu}), f_i(\boldsymbol{\mu})$, and $g_{\mathbf{h}}(\boldsymbol{\mu})$ are affine function of $\boldsymbol{\mu}$ and $O_i(\boldsymbol{\mu})$ is a convex function of $\boldsymbol{\mu}$.
Hence, the reformulated optimization problem \eqref{eq:SRP_new} is non-convex as the left-hand-side of \eqref{eq:SRP_PowerConstraint} is a sum of affine ratios, which is non-convex. Nevertheless, the objective function in \eqref{eq:Obj_fun_11a} and the equality constraint in \eqref{eq:C2_11c} are convex. 

Below, we provide an algorithm that solves \eqref{eq:SRP_new} optimally. 

\begin{theorem}\label{thm:SRP-Optimal}
    Algorithm~\ref{alg:SRP} solves \eqref{eq:SRP_new} optimally.
\end{theorem}
\begin{proof}
    See Appendix A. 
\end{proof}

\begin{algorithm}[t]
\caption{Algorithm to solve \eqref{eq:SRP_new} for a given $\bar{P}$. In the algorithm,  $\psi ( \boldsymbol{\mu}_k, \boldsymbol{\beta}_k, \boldsymbol{\rho}_k ) \triangleq [-d_1(\boldsymbol{\mu}_k)+ \beta_{1,k} f_1(\boldsymbol{\mu}_k), \ldots, - d_N(\boldsymbol{\mu}_k)+ \beta_{N,k} f_N(\boldsymbol{\mu}_k), -\theta_m + \rho_{1,k} f_1(\boldsymbol{\mu}_k), \ldots, -\theta_m + \rho_{N,k} f_N(\boldsymbol{\mu}_k) ]^T$. 
}\label{alg:SRP}
\begin{algorithmic}
\State \textbf{Step $0$:} Initialize $m = 0$,  $\theta_1 = 0$, $\theta_2 \gg 1$, and $\epsilon \ll 1$.
\State \textbf{Step $1$:} Compute $\theta_m = (\theta_1 + \theta_2)/2$. Choose any $\boldsymbol{\mu}_{0}$ that satisfies $g_{\mathbf{h}}(\boldsymbol{\mu}_0) \leq 0, \forall \mathbf{h} \in \mathcal{H}^N.$ Let $k =0$, $\beta_{i,k} = d_i(\boldsymbol{\mu}_k)/ f_i(\boldsymbol{\mu}_k)$ and $\rho_{i,k} = \theta_m/f_i(\boldsymbol{\mu}_k), \forall i \in \mathcal{N}$. 
\State \textbf{Step $2$:} Find the solution $\boldsymbol{\mu}_k = \boldsymbol{\mu}(\boldsymbol{\beta}_k, \boldsymbol{\rho}_k )$ of \eqref{eq:SRP_new3} for given $\theta_m$.
\If{$\psi (\boldsymbol{\mu}_k,  \boldsymbol{\beta}_k, \boldsymbol{\rho}_k ) = \boldsymbol{0}$} 
    \State $\boldsymbol{\mu}_{k, \theta_m} = \boldsymbol{\mu}_{k}$ is a global solution for given $\theta_m$ and go to \textbf{Step $3$}.
\Else
    \State Update $\boldsymbol{\beta}, \boldsymbol{\rho}$ as $\beta_{i, k+1} \gets d_i(\boldsymbol{\mu}_k)/ f_i(\boldsymbol{\mu}_k)$, $\rho_{i, k+1} \gets \theta_m/f_i(\boldsymbol{\mu}_k), \forall i \in \mathcal{N}$. Increment $k$ and go to \textbf{Step $2$}.
\EndIf
\State \textbf{Step $3$:} Compute $P^{\boldsymbol{\mu}_{k, \theta_m}}$ using \eqref{eq:actual_tot_pow}. 
\If{$|P^{\boldsymbol{\mu}_{k, \theta_m}} - \bar{P}| < \epsilon$} 
    \State $\boldsymbol{\mu}_{k, \theta_m}$ is the optimal solution for given $\bar{P}$ and stop the algorithm.
\Else
    \State Update $\theta_1$ and $\theta_2$ as follows: 
        \If{$P^{\boldsymbol{\mu}_{k, \theta_m}} > \bar{P}$} 
            \State $\theta_1 \gets \theta_m$
        \Else
            \State $\theta_2 \gets \theta_m$
        \EndIf
\\Increment $m$ and go to \textbf{Step $1$}.
\EndIf

\end{algorithmic}
\end{algorithm}

As mentioned, we are also interested in obtaining the CO-SRP under the TDMA scheme. Under the TDMA scheme, the CO-SRP can be expressed as follows: In a slot, when the channel power gain is $\mathbf{h}$, the BS transmits to User $i\in \mathcal{N}$ with probability  $\mu_{\mathbf{h}}^{i}$. In this case, the equivalent of \eqref{eq:SRP_new} under the TDMA scheme can be obtained by constraining $\mathcal{W}$ to take the values in $\{\phi, \{1\}, \{2\},\ldots,\{N\}\}$, instead of all the values in the power set $2^{\mathcal{N}}$ and it can be solved via Algorithm \ref{alg:SRP} with appropriate modifications. 

Algorithm \ref{alg:SRP} is an iterative algorithm that obtains the optimal solution to the non-convex problem. The solution in  {Step $2$} can be obtained by standard numerical techniques as for given $\theta, \boldsymbol{\beta}$ and $\boldsymbol{\rho}$, \eqref{eq:SRP_new3} is a convex problem.

\subsubsection{Performance Bound}
The CO-SRP does not consider instantaneous VAoI for making transmission decisions, which may result in suboptimal performance. Nevertheless, we show in the following that the performance of the CO-SRP remains within a bounded multiplicative gap from that of the optimal achievable average VAoI as the solution to \eqref{eq:main-opt-problem}.
\begin{theorem}\label{thm:LB}
The optimal VAoI of the CO-SRP, $V_{\rm SRP}$ in \eqref{eq:SRP_new} is less than or equal to twice the optimal VAoI, $V_{\rm opt}$ in \eqref{eq:main-opt-problem}, i.e., $V_{\rm SRP}\leq 2 V_{\rm opt}$. 
\end{theorem}
\begin{proof}
The proof consists of two parts. First, we obtain a lower bound for the objective function of \eqref{eq:main-opt-problem} using an optimization problem. Second, we compare this lower bound with the objective function of the CO-SRP to prove the theorem. See Appendix B for details. 
\end{proof}

\subsection{CMDP Based Solution}
In this section, we cast the problem \eqref{eq:main-opt-problem} under the CMDP framework. We then relax it using the Lagrangian approach and obtain the optimal solution for the relaxed problem. The parameters and decision variables of a CMDP associated with \eqref{eq:main-opt-problem} are as follows: Let $\mathbf{s} \triangleq (\boldsymbol{\Delta},\mathbf{h})$ be the state of the system, where $\boldsymbol{\Delta} \triangleq (\Delta_1,\ldots, \Delta_N)$ and $\mathbf{h} \triangleq (h_1,\ldots, h_N)$ are the set of VAoIs and set of channel power gain realizations of all users in time slot $t$, respectively. 
The set of actions for all users at a time step is denoted as $\mathbf{u} \triangleq (u_1,\ldots,u_N)$, where $u_i = 0$ indicates that the packet is not transmitted from Queue $i$ and $u_i = 1$ indicates that the packet is successfully transmitted from Queue $i$, which requires sufficient power to be allocated to deliver $R^0_{i}$ number of bits. 
We can compute the probability transition matrix based on the packet arrival and channel power gain distributions for each $\mathbf{s}$ and $\mathbf{u}$. We define a stationary policy, $\pi$, which is a mapping from state space to action space. For a given stationary policy $\pi$, the long-term expected average weighted sum of VAoI and the expected power consumption cost will be:
\begin{align*}
    &{V}_\pi = \lim_{T\rightarrow \infty} \frac{1}{T}\sum_{t=1}^{T}\sum_{i=1}^{N}w_i\mathbb{E}_{\pi}[\Delta_i(t)], \quad \text{and}\\
    &{P}_\pi  = \lim_{T\rightarrow \infty}\frac{1}{T}\sum_{t=1}^{T} \sum_{i=1}^N\mathbb{E}_{\pi}[u_i(t) P_i(t)], 
\end{align*}   
respectively. Now, \eqref{eq:main-opt-problem} can be rewritten as the CMDP problem: 
\begin{subequations}\label{eq:CMDP_opt_problem}
\begin{align}
V_{\rm opt} = \min_{\pi}\;\;&V_\pi,\\
\text{subject to}\;&P_\pi \leq \bar{P}.\label{eq:CMDP_constraint}
\end{align}
\end{subequations}
 To obtain the solution for \eqref{eq:CMDP_opt_problem}, we transform it into an unconstrained MDP problem using the Lagrangian relaxation method, which is discussed below.
\subsubsection{Lagrange Relaxation of the CMDP}
We now relax \eqref{eq:CMDP_opt_problem} to an unconstrained MDP problem using Lagrangian relaxation. The average Lagrangian cost function is defined as
\begin{align*}
    &\mathcal{L}_\pi(\theta) = V_\pi + \theta P_\pi \nonumber \\
    & = \lim_{T \rightarrow \infty} \frac{1}{T} \sum_{t=1}^T \sum_{i=1}^N  \mathbb{E}_{\pi}[w_i \Delta_i(t) + \theta u_i(t) P_i(t)],  
\end{align*}
where $\theta$ is the Lagrange multiplier and the random instantaneous cost is  $C(\mathbf{s},\mathbf{u}, \theta) = \sum_{i=1}^N (w_i \Delta_i + \theta u_i P_i)$. Now, \eqref{eq:CMDP_opt_problem} can be transformed as an unconstrained problem, 
\begin{align}\label{eq:MDP_problem}
    \mathcal{L}^{\rm opt}_{\pi}(\theta) &=\min_{\pi \in \Pi}\; \mathcal{L}_\pi( \theta), 
\end{align}
where $\mathcal{L}^{\rm opt}_\pi(\theta)$ is the minimum average Lagrangian cost obtained by the optimal policy $\pi^{\rm opt}_{\theta}$ for a given $\theta$. We first obtain the optimal policy $\pi^{\rm opt}_{\theta}$ of the unconstrained MDP problem by solving the Bellman equation.

In the optimal case, the solution to \eqref{eq:MDP_problem} will result in a bounded VAoI and bounded instantaneous cost for each User $i\in \mathcal{N}$. This is because if the VAoI of User $i$ becomes greater than $\theta P_i^{\rm max}$, where $P_i^{\rm max}$ is the worst case maximum power required to deliver $R^0_i$ bits of User $i$, we can always transmit the packet (since the VAoI is non-zero, there will be a packet for transmission)  and reduce the overall cost. Hence, in our case, the MDP can be considered to have finite state and action spaces with a bounded instantaneous cost. 
To obtain our results, we consider the following discounted total cost criterion 
\begin{subequations}
\begin{align*}
V^{\theta}_{\gamma}(\mathbf{s}) = \min_{\pi} &\;\lim_{T\rightarrow \infty}\frac{1}{T}\sum_{t=1}^{T}\gamma^{t-1}\sum_{i=1}^{N}\mathbb{E}_{\pi}[w_i\Delta_i(t)+\theta u_i(t) P_i(t)],
\end{align*}
\end{subequations} 
where $0<\gamma<1$ is the discount factor. It can be shown that $V^{\theta}_{\gamma}(\mathbf{s})$ satisfies 
	\begin{align}\label{eq:VI}
	V^{\theta}_\gamma(\mathbf{s}) &= \min_{\mathbf{u}} Q_{\gamma}(\mathbf{s},\mathbf{u}, \theta),
	\end{align}
 where $Q_{\gamma}(\mathbf{s},\mathbf{u}, \theta) =   \mathbb{E}[C(\mathbf{s},\mathbf{u}, \theta) + \gamma V^{\theta}_\gamma(\mathbf{s}')| \mathbf{s, u}]$, where $\mathbf{s}'$ is the next state reached from state $\mathbf{s}$, when action $\mathbf{u}$ is taken, as in Proposition 5 of \cite{Modiano_No_Buffer_Case}. Further, we define, 
\begin{align}\label{eq:val_iteration}
V^{\theta}_{\gamma, m+1}(\mathbf{s}) = \min_{\mathbf{u}}  \; \mathbb{E}[C(\mathbf{s},\mathbf{u}, \theta) + \gamma V^{\theta}_{\gamma,m}(\mathbf{s}')| \mathbf{s, u}],
\end{align}
for all  $m \in \{0,1,2,\ldots\}$ with  $V^{\theta}_{\gamma,0}(\mathbf{s})=0$ for all $\mathbf{s}$, it can be shown that $\lim_{m\to \infty} V^{\theta}_{\gamma, m}(\mathbf{s}) = V^{\theta}_{\gamma}(\mathbf{s})$ for each $\mathbf{s}$ and $\gamma\in (0,1)$ and that $\lim_{\gamma \rightarrow 1} V^{\theta}_{\gamma}(s)$ will give the optimal objective value to \eqref{eq:MDP_problem}.  


Let $\boldsymbol{x}_{-i} \triangleq (x_1,\ldots,x_{i-1}, x_{i+1},\ldots,x_N)$. Below, we show the optimal policy obtained by the MDP is of threshold-type. 
 
\begin{theorem}\label{thm:MDP}
For any $i \in \{1,2,\ldots, N\}$, suppose $\boldsymbol{\Delta_{-i}}$, $\mathbf{h}$ and $\mathbf{u}_{-i}$ are fixed. If it is optimal to transmit $i^{\rm th}$ stream when its VAoI is $\Delta_i$, then it must be optimal to transmit $i^{\rm th}$ stream when its VAoI is $\Delta_i+1$. 
Similarly, for any $i \in \{1,2,\ldots, N\}$, suppose $\boldsymbol{\Delta}$, $\mathbf{h}_{-i}$ and $\mathbf{u}_{-i}$ are fixed, if it is optimal to transmit $i^{\rm th}$ stream when the channel for User $i$ is $h_i$, then it must be optimal to transmit $i^{\rm th}$ stream, when the channel is greater than $h_{i}$.
\end{theorem}
\begin{proof}
    See Appendix C.  
\end{proof}
Due to the above theorem, the complexity of solving and executing the considered problem using the MDP-based approach reduces.

\subsubsection{Bisection Search for $\theta$}
Given the optimal policy $\pi^{\rm opt}_{\theta}$ for \eqref{eq:MDP_problem} for a given $\theta$, 
the objective function of the CMDP problem \eqref{eq:CMDP_opt_problem} increases with $\theta$, and $P_{\pi^{\rm opt}_{\theta}}$ in \eqref{eq:CMDP_constraint} decreases with $\theta$. To determine a suitable value of $\theta$ that makes the result of the unconstrained problem to satisfy the constraint \eqref{eq:CMDP_constraint} with equality, we employ the bisection algorithm.\\

\section{Numerical Results}\label{sec:num_results}

In this section, we present numerical results for the two-user case. We consider $f(x) = \log(1+x)$.

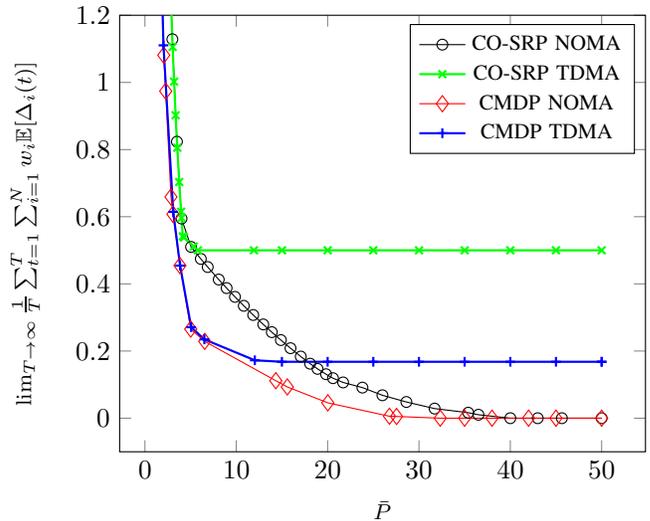
\begin{figure}[t]
    \centering
    \begin{tikzpicture}[scale=1.02]
      \begin{axis}[
        xlabel={$\bar{P}$},
        ylabel={$\lim_{T\rightarrow \infty} \frac{1}{T}\sum_{t=1}^{T}\sum_{i=1}^{N}w_i\mathbb{E}[\Delta_i(t)]$},
        ymax=1.2,
        ymin=-0.1,
        every axis label/.append style={align=center, font=\footnotesize},
        legend style={font=\footnotesize}
      ]

      \pgfplotstableread[col sep=comma]{Results/pow_vs_age_srp_noma.csv}\datatable 
      \addplot [color=black, mark=o, mark options={solid, black}, mark size = 2pt ]table[x=pow,y=age_noma]{\datatable};
      \addlegendentry{CO-SRP NOMA};

      \pgfplotstableread[col sep=comma]{Results/pow_vs_age_srp_tdma.csv}\datatable 
      \addplot [color=green, mark=x, mark options={solid, green}, mark size = 2pt, line width=0.8pt]table[x=pow,y=age_tdma]{\datatable};
      \addlegendentry{CO-SRP TDMA};

      \pgfplotstableread[col sep=comma]{Results/pow_vs_age_MDP_6.csv}\datatable 
      \addplot [color= red, mark=diamond, mark options={solid, red}, mark size = 3pt]table[x=MDP_pow,y=MDP_age]{\datatable};
      \addlegendentry{CMDP NOMA};

      \pgfplotstableread[col sep=comma]{Results/pow_vs_age_MDP_6_tdma.csv}\datatable 
      \addplot [color= blue, mark=+, mark options={solid, blue}, mark size = 1.8pt, line width=0.8pt]table[x=MDP_pow,y=MDP_age]{\datatable};
      \addlegendentry{CMDP TDMA};

      \end{axis}
    \end{tikzpicture}

    \caption{Variation of the long-term expected average weighted sum of VAoI, with the average transmit power, $\bar{P}$, under the CO-SRP and the CMDP-based policy with NOMA and TDMA schemes. Parameters adopted are $\lambda_1 = \lambda_2 = 0.5$, $R^0_1=R^0_2=2$, $\mathcal{H}\in \{1, 0.1\}$, $\mathbb{P}(H_1=0.1) = \mathbb{P}(H_2=0.1) =0.2$, and $w_1 = w_2 = 0.5$.}
    \label{fig:pow_vs_age}
\end{figure}

In Fig.~\ref{fig:pow_vs_age}, we show the variation of the long-term expected average VAoI as a function of the average transmit power, $\bar{P}$,  for NOMA and TDMA schemes under CO-SRP and CMDP-based policy. 
 Firstly, we observe that as $\bar{P}$ increases, the average VAoI decreases. This is because more and more of the arriving packets get delivered as soon as they arrive. Interestingly, at lower powers, NOMA and TDMA perform the same. NOMA starts performing better than TDMA when power increases. This is because, even in NOMA, in the optimal case, packets are not transmitted to both users simultaneously at lower powers, thereby making it similar to TDMA. However, at higher powers, it is optimal for NOMA to transmit to both users simultaneously. Secondly, we observe that the average VAoI tends to zero in the NOMA case as $\bar{P}$ increases, but for the TDMA scheme, it reaches a saturation point at a value of $\lambda$. This behavior can be explained as follows. The expression for the long-term expected average weighted sum of VAoI under CO-SRP is  $\sum_{i=1}^{N} w_i (\lambda_i( p_{i}^{-1} - 1 ))$, where $p_i \in [0, 1]$ is the probability of successful delivery of a packet to User $i$. For the two-user case, we can rewrite expression as $ 0.5 \lambda  ( p_{1}^{-1} + p_{2}^{-1} - 2 )$ when $w_1 = w_2 = 0.5$ and $\lambda_1 = \lambda_2 = \lambda$. With the TDMA scheme, even with infinite available power, the BS can transmit a packet to only one user at any time $t$; this implies $p_2 = (1 - p_1 )$. When $p_1 \to 1$ and $p_2 \to 0$, the average VAoI for the first user goes to zero, while for the second user, it tends towards $\infty$. So, the optimal value for the probability of successful delivery with infinite $\bar{P}$, by symmetry, is $p_1 = p_2 = 0.5$. From this, we conclude that the minimum achievable average VAoI in the TDMA scheme for the two-user case, at high $\bar{P}$, is $\lambda$. However, in the NOMA scheme, there is no restriction on selecting only one user to transmit at any given time. With higher available power, the BS can simultaneously transmit to multiple users. So, the optimal value for the probability of successful delivery of a packet for User $i$ in the NOMA scheme is  $p_i = 1$. Using this condition, we conclude that the minimum achievable average VAoI in the NOMA scheme is zero. In the CMDP-based solution, the trends observed in the CO-SRP continue to hold. Accordingly, in the rest of the section, we obtain results only for CO-SRP.


\begin{figure}[t]
    \centering
    \begin{tikzpicture}[scale=1.03]
      \begin{axis}[
        xlabel={$\bar{P}$},
        ylabel={$\mathbb{E_{\mathbf{H}}}\left[\mu_{\mathbf{h}}^{{\mathcal{W } = \{1,2\} }}\right]$},
        every axis label/.append style={align=center, font=\footnotesize},
        legend style={font=\footnotesize},
        legend pos = north west,
        xmax= 52
      ]

      \pgfplotstableread[col sep=comma]{Results/pow_vs_prob_vals_0_5.csv}\datatable 
      \addplot [color=green, mark= diamond, mark options={green}, mark size = 2pt]table[x=pow,y=prob_vals, restrict x to domain=0:50]{\datatable};
      \addlegendentry{$\mathbb{P}(H=0.1) = 0.5$};

      \pgfplotstableread[col sep=comma]{Results/pow_vs_prob_vals_0_2.csv}\datatable 
      \addplot [color=red, mark=+, mark options={solid, red}, mark size = 2.5pt]table[x=pow,y=prob_vals, restrict x to domain=0:50]{\datatable};
      \addlegendentry{$\mathbb{P}(H =0.1) =0.2$};
    
      \pgfplotstableread[col sep=comma]{Results/pow_vs_prob_vals_0_1.csv}\datatable 
      \addplot [color=blue, mark=o, mark options={solid, blue}, mark size = 1.6pt]table[x=pow,y=prob_vals]{\datatable};
      \addlegendentry{$\mathbb{P}(H =0.1) =0.1$};
   
      \end{axis}
    \end{tikzpicture}
    \caption{Probability of simultaneous transmissions to both users versus the average transmit power, $\bar{P}$, under the CO-SRP with NOMA scheme for $R^0_1 = R^0_2 = 2$, $\mathcal{H}\in \{1, 0.1\}$, $H_1 = H_2 = H$, $\lambda_1 = \lambda_2 = 0.8$ and $w_1 = w_2= 0.5$.}
    \label{fig:pow_vs_tr_prob}
\end{figure}
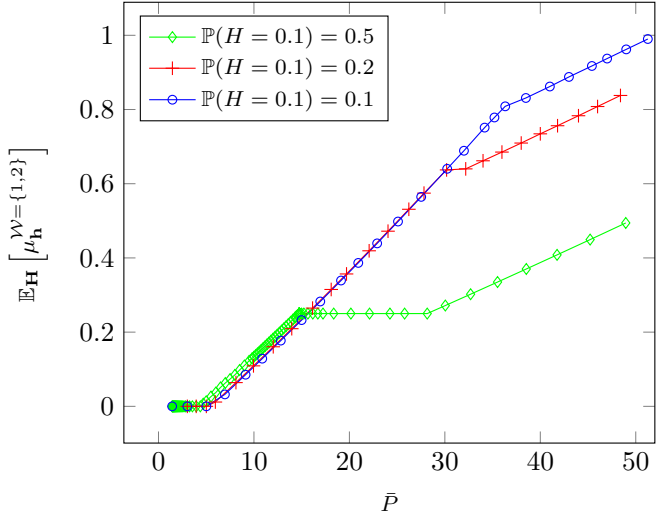

We study the variation of the probability of simultaneous transmission to both users, ($\mathbb{E_{\mathbf{H}}}\left[\mu_{\mathbf{h}}^{{\mathcal{W } = \{1,2\} }}\right]$), in the NOMA scheme as a function of average transmit power, in Fig.~\ref{fig:pow_vs_tr_prob}. From the figure, we observe that at lower  $\bar{P}$, the probability of simultaneous transmission to both users is zero, and hence NOMA behaves exactly like TDMA. However, this probability increases as $\bar{P}$ rises. We also examined this behavior under various scenarios involving the probability of encountering a bad channel (channel with lower channel power gain, $H = 0.1$). When $\mathbb{P}(H=0.1)$ is higher, the power required for transmitting to both users is also high, so NOMA transmits to a single user at any given time with less available power rather than attempting simultaneous transmission to both users. 

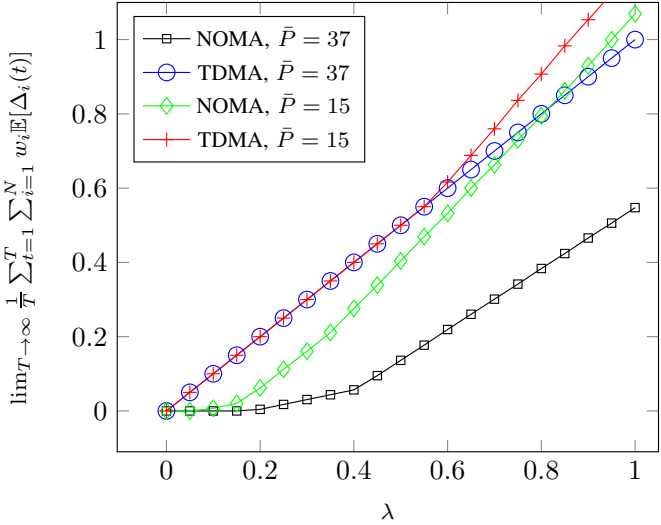
\begin{figure}[t]
    \begin{tikzpicture}[scale=1.05]
      \begin{axis}[
        xlabel={$\lambda$},
        ylabel={$\lim_{T\rightarrow \infty} \frac{1}{T}\sum_{t=1}^{T}\sum_{i=1}^{N}w_i\mathbb{E}[\Delta_i(t)]$},
        every axis label/.append style={align=center, font=\footnotesize},
         legend style={font=\footnotesize},
        legend pos = north west,
        xmax = 1.05,
        ymax = 1.1
      ]

      \pgfplotstableread[col sep=comma]{Results/lambda_vs_age_noma_37.csv}\datatable 
      \addplot [color=black, mark= square, mark options={black}, mark size = 1.5pt]table[x=lambda,y=age_noma]{\datatable}; 
       \addlegendentry{NOMA, $\bar{P} = 37$};

      \pgfplotstableread[col sep=comma]{Results/lambda_vs_age_tdma_37.csv}\datatable 
      \addplot [color=blue, mark= o, mark options={blue}, mark size = 3pt]table[x=lambda,y=age_tdma]{\datatable}; 
       \addlegendentry{TDMA, $\bar{P} = 37$};

       \pgfplotstableread[col sep=comma]{Results/lambda_vs_age_noma_15.csv}\datatable 
      \addplot [color=green, mark= diamond, mark options={green}, mark size = 3pt]table[x=lambda,y=age_noma]{\datatable}; 
       \addlegendentry{NOMA, $\bar{P} = 15$};

     \pgfplotstableread[col sep=comma]{Results/lambda_vs_age_tdma_15.csv}\datatable 
      \addplot [color=red, mark= +, mark options={red}, mark size = 2.5pt]table[x=lambda,y=age_tdma]{\datatable}; 
       \addlegendentry{TDMA, $\bar{P} = 15$};

       \end{axis}
    \end{tikzpicture}

    \caption{Variation of the long-term expected average weighted sum of VAoI with respect to probability of packet arrival, $\lambda_1 = \lambda_2 = \lambda$, under the CO-SRP with NOMA and TDMA schemes. Parameters are set to $R^0_1= R^0_2 = 2$, $\mathcal{H}\in \{1, 0.1\}$, $\mathbb{P}(H_1=0.1) = \mathbb{P}(H_2=0.1) =0.5$, $w_1 = w_2 = 0.5$.}
    \label{fig:lambda_vs_age}
\end{figure}

 In Fig.~\ref{fig:lambda_vs_age}, we show the variation of the long-term expected average VAoI with respect to the probability of packet arrival, $\lambda$, for both NOMA and TDMA schemes under the CO-SRP. NOMA achieves a lower average VAoI compared to the TDMA scheme. 

\begin{figure}[t]
    \begin{tikzpicture}[scale=1.03]
      \begin{axis}[
        xlabel={$\lim_{T\rightarrow \infty} \frac{1}{T}\sum_{t=1}^{T} w_1\mathbb{E}[\Delta_1(t)]$},
        ylabel={$\lim_{T\rightarrow \infty} \frac{1}{T}\sum_{t=1}^{T} w_2\mathbb{E}[\Delta_2(t)]$},
        every axis label/.append style={align=center, font=\footnotesize},
         legend style={font=\footnotesize},
        legend pos = north east,
        ymin=-0.5,
        ymax=5,
        xmin=-0.5,
        xmax=5,
      ]
    
       \pgfplotstableread[col sep=comma]{Results/age_u1_vs_age_u2_weights.csv}\datatable 
      \addplot [name path=noma, color=blue, very thick,  mark options={blue}, mark size = 1.8pt]table[x=age_u1_noma,y=age_u2_noma]{\datatable}; 
       \addlegendentry{NOMA }; 

       \pgfplotstableread[col sep=comma]{Results/age_u1_vs_age_u2_weights.csv}\datatable 
      \addplot [dashed, name path=tdma, color=red, ultra thick, mark options={red, dashed}, mark size = 1.8pt]table[x=age_u1_tdma,y=age_u2_tdma]{\datatable}; 
       \addlegendentry{TDMA };

     \pgfplotstableread[col sep=comma]{Results/age_u1_vs_age_u2_weights.csv}\datatable 
      \addplot [name path=out_curve, color=black, mark options={red}, mark size = 1.8pt]table[x=test1,y=test2]{\datatable}; 
    
       \addplot [blue!10] fill between [of = tdma and noma];
       \addplot [red!5, pattern=north east lines, pattern color=red] fill between [of = out_curve and tdma];
  
       \end{axis}
    \end{tikzpicture}

    \caption{The largest achievable long-term expected average VAoI region by NOMA and TDMA schemes under CO-SRP by varying the weights associated with users ($w_1, w_2$). Parameters adopted are $\bar{P} = 40$, $\lambda_1 = \lambda_2 = 0.9$, $R^0_1 = R^0_2 = 2$ and  $\mathcal{H}\in \{1, 0.1\}$, and $\mathbb{P}(H_1=0.1) = \mathbb{P}(H_2=0.1) =0.5$.}
    \label{fig: bound_on_VAoI}
\end{figure}
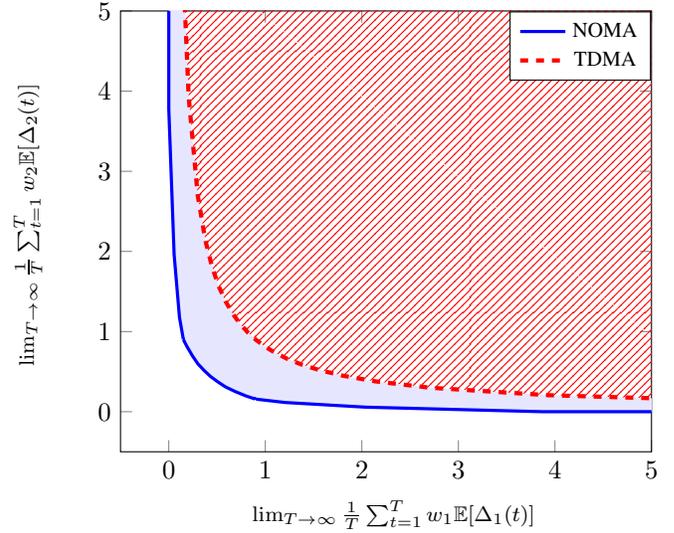

In Fig.~\ref{fig: bound_on_VAoI}, we show the largest achievable long-term expected average VAoI region for both TDMA and NOMA schemes. We varied the weights associated with users ($w_1, w_2$) between $0$ and $1$, subject to the constraint $w_1 + w_2 = 1$. The figure shows that the achievable region of the NOMA scheme includes that of the TDMA scheme. In TDMA, if the average VAoI of any of one user is low, it results in a significantly higher average VAoI for the other user because TDMA allows the BS to transmit to only one user at a given time. 
 
 We also study the variation of the long-term expected average VAoI for TDMA and NOMA schemes with respect to different parameters. The results are summarized in Table~\ref{tab:A_srp_vs_other_parameters}. At higher values of $\bar{P}$, we observe a decrease in both the average VAoI under the TDMA  scheme and the NOMA scheme as the probability of bad channel occurrence, $\mathbb{P}(H=0.1)$, decreases. The average VAoI under NOMA decreases with increasing $H_{\rm max}$ (the maximum value of channel power gain) and decreasing $R^0$. However, these changes in  $H_{\rm max}$ and $R^0$ do not significantly impact the average VAoI under TDMA at higher values of  $\bar{P}$ and $\mathbb{P}(H=0.1)$. 


\begin{table}[t]										
	\centering									
	\caption{Variation of difference of long-term expected average  weighted sum of VAoI under CO-SRP with NOMA and TDMA schemes with respect to $H_{\rm max}$ (maximum value of channel power gain),  $\mathbb{P}(H= 0.1)$ and $R^0$ considering $R^0_1=R^0_2 = R^0$, $H_1 = H_2 = H$, $\mathcal{H}\in \{H_{\rm max}, 0.1 \}$, $\lambda_1 = \lambda_2 = 0.9$, and $w_1 = w_2 = 0.5$. We denote $\bar{\Delta}_{\rm TDMA}$, and $\bar{\Delta}_{\rm NOMA}$ as the expected average VAoI under TDMA and NOMA schemes respectively.}									
	\begin{tabular}{|p{0.6cm}|p{1cm}|p{0.4cm}| p{0.4cm}|p{1cm}|p{1cm}|p{1.3cm}|}									
		\hline
          \vspace{0.05cm}
		 $H_{\rm max}$ & \vspace{0.05cm} $\mathbb{P}(H = 0.1)$ & \vspace{0.05cm} $R^0$ & \vspace{0.05cm} $\bar{P}$ & \vspace{0.05cm}$\bar{\Delta}_{\rm TDMA}$ & \vspace{0.05cm}$\bar{\Delta}_{\rm NOMA}$ &\vspace{0.05cm} $\bar{\Delta}_{\rm TDMA} -\bar{\Delta}_{\rm NOMA}$     \\								
		\hline								
	      $1$  &  $0.9$ & $2$ & $45$ & $1.0774$ & $1.0668$ & $0.0106$ \\		 
            \hline								
		  $1$  &  $0.5$ & $2$ & $45$ & $0.9$ & $0.3850$ & $0.515$ \\	
            \hline								
		$2$  &  $0.5$ & $2$ & $45$ &  $0.9$ & $0.2681$ & $0.6319$\\		
		\hline								
		$2$ & $0.5$ & $3$ & $45$ &  $0.9751$ & $0.9751$ & $0$  \\		
            \hline								
		$2$ & $0.5$ & $3$ & $100$ & $0.9751$ & $0.5265$ & $0.4486$ \\	
            \hline								
		
	\end{tabular}									
	\label{tab:A_srp_vs_other_parameters}									
\end{table}

\section{Conclusions}\label{sec:conclusions}
In this work, we considered the problem of scheduling the transmission of version update packets from exogenous streams to different users via a fading broadcast channel using the NOMA scheme to minimize the long-term expected average weighted sum of VAoI across the users subject an average power constraint at the BS. To address this problem, we derived a simpler, sub-optimal, channel-only stationary randomized policy by solving a non-convex problem optimally and obtained its performance bounds. We also obtained a CMDP-based solution and characterized its structural properties. 
Through numerical simulations, we conclude that our proposed CO-SRP solution achieves competitive performance with the CMDP-based approach while being significantly simpler, and
the VAoI achieved by CO-SRP is within twice the optimal VAoI. Additionally, based on our comparison of NOMA to a TDMA scheme, we conclude that NOMA outperforms TDMA when bound on the power consumption is high. However, NOMA and TDMA exhibit similar performance when this bound low.

 
\bibliographystyle{ieeetr}
\bibliography{references}

\appendix

\subsection{Proof of Theorem~\ref{thm:SRP-Optimal}}
To solve the problem \eqref{eq:SRP_new}, we first consider, 
\begin{subequations}\label{eq:SRP_new1}
\begin{align}
\underset{{0 \leq \mu_{\mathbf{h}}^{\mathcal{W}} \leq 1}}{\text{minimize}}\;\; &\sum_{i=1}^N w_i O_i(\boldsymbol{\mu}) + \theta \left(\;\sum_{i=1}^N\frac{d_i(\boldsymbol{\mu})}{f_i(\boldsymbol{\mu})}-\bar{P}\right),\\
\text{subject to}\label{eq:constraint}\;\;
&g_{\mathbf{h}}(\boldsymbol{\mu}) = 0, \quad \forall \mathbf{h} \in \mathcal{H}^N,
\end{align}
\end{subequations}
where $\theta \geq 0$ is the Lagrange multiplier corresponding to the average power constraint, \eqref{eq:SRP_PowerConstraint}.  
For a given $\theta$,  \eqref{eq:SRP_new1} is a sum-of-ratios problem. 
Now, we introduce an auxiliary variable $\beta_i$ and obtain the following problem equivalent to \eqref{eq:SRP_new1}:  
\begin{subequations}\label{eq:SRP_new2}
\begin{align}
\underset{0 \leq \mu_{\mathbf{h}}^{\mathcal{W}} \leq 1, \beta_i}{\text{minimize}} &\sum_{i=1}^N w_i O_i(\boldsymbol{\mu}) + \theta\left(\sum_{i=1}^N \beta_i - \bar{P}\right),\\
\text{subject to}\;\;
& d_i(\boldsymbol{\mu}) -  \beta_i f_i(\boldsymbol{\mu})\leq 0, \; \forall i\in \mathcal{N},\label{eq:c1}\\
&g_{\mathbf{h}}(\boldsymbol{\mu}) = 0, \quad \forall \mathbf{h} \in \mathcal{H}^N. \label{eq:c2}
\end{align}
\end{subequations}
\eqref{eq:SRP_new1} and \eqref{eq:SRP_new2} are equivalent because of the following:  if $\boldsymbol{\mu}$ is optimal solution to \eqref{eq:SRP_new1}, then $(\boldsymbol{\mu}, \boldsymbol{\beta})$ is optimal solution for \eqref{eq:SRP_new2}, where $\beta_i = {d}_i(\boldsymbol{\mu})/f_i(\boldsymbol{\mu})$. 
Similarly, if $(\boldsymbol{\beta}, \boldsymbol{\mu})$ is optimal solution to \eqref{eq:SRP_new2},  then $\boldsymbol{\mu}$ be the optimal solution to \eqref{eq:SRP_new1}. In the optimal case $\beta_i = d_i(\boldsymbol{\mu})/f_i(\boldsymbol{\mu}), \forall i \in \mathcal{N}$, as if not we can always reduce $\beta_i$ to equal to this and reduce the objective function.

For a given $\theta$, \eqref{eq:SRP_new2} is a non-convex problem due to the product of $\beta_i$ and $f_i(\boldsymbol{\mu})$ in \eqref{eq:c1}. Let $\rho_i$ and $v_{\mathbf{h}}$ be the dual variables corresponding to \eqref{eq:c1} and \eqref{eq:c2}, respectively. The Lagrange function of \eqref{eq:SRP_new2} is given by
    \begin{align*}
        &L(\boldsymbol{\mu}, \boldsymbol{\beta}, \boldsymbol{\rho}, \boldsymbol{v}) =  \sum_{i=1}^N w_i O_i(\boldsymbol{\mu}) + \theta\left(\sum_{i=1}^N \beta_i-\bar{P}\right)\nonumber\\
        &+ \sum_{i=1}^N \rho_i \left({d}_i(\boldsymbol{\mu}) - \beta_i f_i(\boldsymbol{\mu}) \right) + \sum_{ \mathbf{h} \in \mathcal{H}^N} v_{\mathbf{h}}\; g_{\mathbf{h}}(\boldsymbol{\mu}),
    \end{align*}
where $\boldsymbol{\beta} \triangleq \{\beta_i, \ldots, \beta_N \}$, $\boldsymbol{\rho} \triangleq \{\rho_i, \ldots, \rho_N \}$ and $\boldsymbol{v} = \{v_{\mathbf{h}} |\; \forall \mathbf{h} \in \mathcal{H}^N \}$.
Although \eqref{eq:SRP_new2} is non-conex, the  Karush-Kuhn-Tucker (KKT) conditions are necessary for optimality \cite{Boyd}. Specifically, among other KKT conditions, the optimal solution must satisfy:
\begin{align*}
\frac{\partial L(\boldsymbol{\mu}, \boldsymbol{\beta}, \boldsymbol{\rho}, \boldsymbol{v})}{\partial \beta_i}  =  \theta - \rho_i f_i(\mathbf{\mu}) &= 0, \forall i \in \mathcal{N}, \text{and}\\
\rho_i \left({d}_i(\boldsymbol{\mu}) - \beta_i f_i(\boldsymbol{\mu}) \right) & = 0, \forall i \in \mathcal{N}.
\end{align*}
From the KKT conditions, note that for any given $\theta>0$,  $\rho_i$ must be greater than zero because $f_i(\boldsymbol{\mu})>0$ for any $\boldsymbol{\mu}$. Hence, we have,
\begin{align}
    d_i(\boldsymbol{\mu}) -  \beta_i f_i(\boldsymbol{\mu}) &= 0, \;\forall i\in \mathcal{N}, \text{and}\label{eq:beta}\\
    \theta - \rho_i f_i(\boldsymbol{\mu}) &= 0, \;\forall i\in \mathcal{N}. \label{eq:gamma}
\end{align}
Let $\Omega = \{\alpha = (\boldsymbol{\beta}, \boldsymbol{\rho}):\eqref{eq:beta}, \eqref{eq:gamma}\}$. If $\Omega$ contains a single element, it must be optimal, and optimal $\boldsymbol{\mu}$ can be obtained by solving 
\begin{subequations}\label{eq:SRP_new3}
\begin{align}
\underset{0 \leq \mu_{\mathbf{h}}^{\mathcal{W}} \leq 1}{\text{minimize}} &\sum_{i=1}^N w_i O_i(\boldsymbol{\mu}) + \theta\left(\sum_{i=1}^N \beta_i - \bar{P}\right) \nonumber\\
&+ \sum_{i=1}^N\rho_i \left(d_i(\boldsymbol{\mu}) -  \beta_i f_i(\boldsymbol{\mu})\right)\\
\text{subject to} 
&\;\; g_{\mathbf{h}}(\boldsymbol{\mu}) = 0, \quad \forall \mathbf{h} \in \mathcal{H}^N. \label{eq:c22}
\end{align}
\end{subequations}

Let $\boldsymbol{\mu}(\alpha)$ be the solution to \eqref{eq:SRP_new3} for a given $\alpha$. Since for any feasible $\boldsymbol{\mu}$ to \eqref{eq:SRP_new1}, we can get $\alpha$ from \eqref{eq:beta} and \eqref{eq:gamma}, and we can get $\alpha^{\rm opt}$ such that $\boldsymbol{\mu}^{\rm opt} = \boldsymbol{\mu}(\alpha^{\rm opt})$. Now, for a given $\theta$, the uniqueness of  $\alpha^{\rm opt}$ following from Theorem~3.1 and Corollary 3.1 in \cite{jong2012efficient} and an algorithm to find it can be obtained via Algorithm~\ref{alg:SRP}, similar to the modified Newton algorithm proposed in \cite{jong2012efficient}.
We employ Algorithm~\ref{alg:SRP} to iteratively solve \eqref{eq:SRP_new3} in order to find the optimal solution to \eqref{eq:SRP_new} for a given $\theta \geq 0$. Additionally, we utilize the bisection method to search for suitable value of $\theta$ that satisfy the average power constraint, \eqref{eq:SRP_PowerConstraint}, with equality.

\subsection{Proof of Theorem~\ref{thm:LB}}

We first derive a lower-bound on the long-term expected average sum of  VAoI $\lim_{T\rightarrow \infty}({1}/{T})\sum_{t=1}^{T}\sum_{i=1}^{N}w_i\mathbb{E}[\Delta_i(t)]$. 
For a given sample path of packet arrival and delivery over the time horizon of $T$ units, let $D_i(T)$ be the total number of packets delivered to the Destination $i$ until time slot $T$. $t_i(m)$ be the time slot of $m^{\rm th}$ delivery to the Destination $i$, $\forall m \in \{1,\ldots, D_i(T)\}$. We define  inter-delivery time as {$I_i(m):= t_i(m)$ when $m = 1$, and $I_i(m):= t_i(m) - t_i(m-1)$ for all $m > 1$, i.e., the number of time slots between $(m-1)^{\rm th}$ and  $m^{\rm th}$ deliveries to the destination. We note that after every inter-delivery time, $I$, which is a random variable, the VAoI of a user gets reset to zero. 
We now characterize the expected sum of VAoI in an inter-delivery time. For brevity, we drop user and packet indices where not required. 
Let $n$ be the number of packet arrivals during an inter-delivery time of duration $I$ slots.  
For a given $I$, the number of arrivals, $n\in \{1,2,\ldots,I\}$, as, at most one packet can arrive in a slot. 
When $n=1$, a packet arrival can happen in any one of the positions in $\{1,\ldots,I\}$. If an arrival occurs at the beginning of Slot $1$, the total  VAoI summed over $I$ slots is equal to $I-1$, and the probability of arrival of single packet exactly in position of Slot $1$ is $1/{}^{I}C_{1}$.  Similarly, if the arrival happens at the beginning of Slot $2$, the total VAoI sum is $I-2$, and the probability of arrival of single packet exactly in position of Slot $2$ is $1/{}^{I}C_{1}$ and so on. Hence, conditioned on $n=1$, the expected sum of VAoI for given $I$ can be expressed as $\mathbb{E}\left[\Delta_I\mid n=1\right]  =(I-1)(1/{}^{I}C_{1}) + (I - 2)(1/{}^{I}C_{1})+\ldots+ 1 \times (1/{}^{I}C_{1}) + 0 \times (1/{}^{I}C_{1}) = (I(I-1)/2)(1/{}^{I}C_{1}) = (I-1)/2$. Similarly, consider the case when $n=2$, where two packet arrivals can occur in any two of the slots in $\{1,\ldots,I\}$, in ${}^{I}C_{2}$ possible ways. For example, when the first arrival occurs at the beginning of Slot $1$ and the second arrival happens at the beginning of Slot $2$,  the sum of VAoI is $(I-1) + (I-2)$ and the probability of arrival of two packets exactly in the positions of Slot $1$ and $2$ is $1/{}^{I}C_{2}$. Similarly, when the first arrival occurs at the beginning of Slot $1$ and the second arrival happens at the beginning of Slot $3$, the sum VAoI is $(I-1) + (I-3)$ and the probability of arrival of two packets exactly in the positions of Slot $1$ and $3$ is $1/{}^{I}C_{2}$, and so on (see Fig.~\ref{fig:illustration_to_measure_VAoI}).    
Hence, conditioned on $n=2$, the expected sum of VAoI for given $I$ can be expressed as  $\mathbb{E}\left[\Delta_I\mid n=2\right]=  ((I-1) + (I-2)) (1/{}^{I}C_{2})  + ((I-1) + (I-3)) (1/{}^{I}C_{2}) +\ldots  + ((I-1) + 0) (1/{}^{I}C_{2})  + ((I -2)+ (I-3)) (1/{}^{I}C_{2}) + ((I-2) + (I-4)) (1/{}^{I}C_{2}) +\ldots  + ((I-2) + 0) (1/{}^{I}C_{2}) 
+\ldots+ (1+0) (1/{}^{I}C_{2}) = ({I(I-1)}/{2}) \times (2/I)$. In general, it can be shown that the expected sum of VAoI conditioned on any given $n\in \{1,2,\ldots,I\}$ can be expressed as  $\mathbb{E}\left[\Delta_I\mid n\right] = ((I-1)/2) \times n$. The expectations are  with respect to the positions of the packet arrival during the inter-delivery time of $I$ slots. Now, we remove the conditioning on $n$. By considering the probability of $n$ number of packet arrivals during the inter-delivery time $I$ as ${}^{I}C_{n} \lambda^n (1-\lambda)^{I-n}$, the expected sum of VAoI for given $I$ can be expressed as
\begin{align*}
    \mathbb{E}[\Delta_I] = \frac{(I-1)}{2}\sum_{n = 1}^{I} n \times {}^{I}C_{n} \times \lambda^{n} (1 - \lambda)^{I - n}.  
\end{align*}
Using  $\sum_{n = 1}^{I} n \; {}^{I}C_{n} \; \lambda^{n} (1 - \lambda)^{I - n} = I\lambda$, in the above equation, gives
\begin{align}\label{eq:expected_delta}
\mathbb{E}[\Delta_I] = \lambda (I^2 - I)/2. 
\end{align}
\begin{figure}[t]
    \centering
\begin{tikzpicture}

    \def\markheight{0.05} 

    \def \m{0}
    \def\xmin{0}
    \def\xmax{5.5}
    \def\ymin{\m}
    \def\ymax{\m+1}

    \draw[->, black, thick] (3.5, \m++0.5+0.3+1.5) -- (3.5, \m+0.5+0.3+2) {};
    \draw[->, black, thick] (2.5, \m+0.5+0.3+ 1.5) -- (2.5, \m+0.5+0.3+2) {};
    \draw[->, black, thick] (4.5, \m+0.5++0.3+ 1.5) -- (4.5, \m+0.5+0.3+2) {};

    \draw[-, black, thick] (2.5, \m+0.5) -- (3.5,\m+0.5) {};
    \draw[-, black, thick] (2.5, \m) -- (2.5,\m+0.5) {};
    \draw[-, black, thick] (3.5,\m+1) -- (3.5,\m+0.5) {};
    \draw[-, black, thick] (3.5,\m+1) -- (4.5,\m+1) {};
    \draw[-, black, thick] (4.5,\m+1) -- (4.5,\m+1+0.5) {};
    \draw[-, black, thick] (4.5,\m+1+0.5) -- (5,\m+1+0.5) {};
    \draw[-, black, thick] (5,\m+1+0.5) -- (5,\m+0) {};

    \draw[dashed, gray ] (3.5, \m+0.5) -- (5,\m+0.5) {};
    \draw[dashed, gray ] (4.5, \m+0.5+0.5) -- (5,\m+0.5+0.5) {};

    \draw[-, black] (\xmin,\m) -- (\xmax,\m) node[right] {};
    \draw[->, black] (\xmin, \m+0.5+0.3+ 1.5) -- (\xmax, \m+0.5+0.3+ 1.5) node[right] {$t$};

    \draw[->, black] (0,\ymin) -- (0,\ymax+0.5+0.3) node[above] {$\Delta$};

    \node at (3.7, \m+0.25) [font=\small] {$2$};
    \node at (4, \m+0.75) [font=\small] {$1$};
    \node at (4.7, \m+1.25) [font=\small] {$0$};

    \foreach \x/\label  in { 0.5/{1}, 1.7/{\ldots}, 3/{$I-2$},  4/{$I-1$}, 5/{$I$}}
        \draw (\x,\m) node[below] [font=\tiny] { \label};

    \foreach \y/\label in { 0.5/{1}, 1/{2}, 1.5/{3}}
        \draw (-0.2,\y) node[left] [font=\tiny] {\label};

    \foreach \x in {0,1,2.5,3.5,4.5, 5.5}
        \draw (\x,\m+\markheight) -- (\x,\m-\markheight);

    \foreach \y in {0.5, 1, 1.5}
        \draw (-\markheight, \y) -- (\markheight, \y);


    \def \m{3.3}
    \def\xmin{0}
    \def\xmax{5.5}
    \def\ymin{\m}
    \def\ymax{\m+1}

    \draw[->, black, thick] (0, \m+1.5+0.3) -- (0, \m+2+0.3) {};
    \draw[->, black, thick] (2, \m+ 1.5+0.3) -- (2, \m+2+0.3) {};

    \draw[-, black, thick] (0, \m+0.5) -- (2,\m+0.5) {};
    \draw[-, black, thick] (2,\m+1) -- (2,\m+0.5) {};
    \draw[-, black, thick] (2,\m+1) -- (5,\m+1) {};
    \draw[-, black, thick] (5,\m+1) -- (5,\m+0) {};

    \draw[dashed, gray ] (2, \m+0.5) -- (5,\m+0.5) {};

    \draw[-, black] (\xmin,\m) -- (\xmax,\m) node[right] {};
    \draw[->, black] (\xmin, \m+ 1.5+0.3) -- (\xmax, \m+ 1.5+0.3) node[right] {$t$};

    \draw[->, black] (0,\ymin) -- (0,\ymax+0.3) node[above] {$\Delta$};

    \node at (2.5, \m+0.25) [font=\small] {$I-1$};
    \node at (3.5, \m+0.75) [font=\small] {$I-3$};

    \foreach \x/\label  in { 0.5/{1}, 1.5/{2}, 2.5/{3}, 3.7/{\ldots},  5/{$I$}}
        \draw (\x,\m) node[below] [font=\tiny] { \label};

    \foreach \x in {0,1,2,3,4.5, 5.5}
    \draw (\x,\m+\markheight) -- (\x,\m-\markheight);

    \foreach \y/\label  in { \m+0.5/{1}, \m+1/{2}}
        \draw (-0.2,\y) node[left] [font=\tiny] { \label};

    \foreach \y in {\m+0.5, \m+1}
        \draw (-\markheight, \y) -- (\markheight, \y);


    \def \m{6.1}
    \def\xmin{0}
    \def\xmax{5.5}
    \def\ymin{\m}
    \def\ymax{\m+1}

    \draw[->, black, thick] (0, \m+1.5+0.3) -- (0, \m+2+0.3) {};
    \draw[->, black, thick] (1, \m+ 1.5+0.3) -- (1, \m+2+0.3) {};

    \draw[-, black, thick] (0, \m+0.5) -- (1,\m+0.5) {};
    \draw[-, black, thick] (1,\m+1) -- (1,\m+0.5) {};
    \draw[-, black, thick] (1,\m+1) -- (5,\m+1) {};
    \draw[-, black, thick] (5,\m+1) -- (5,\m+0) {};

    \draw[dashed, gray ] (1, \m+0.5) -- (5,\m+0.5) {};
    \draw[dashed, gray ] (4.5, \m++0.5+ 0.5) -- (5,\m++0.5+ 0.5) {};

    \draw[-, black] (\xmin,\m) -- (\xmax,\m) node[right] {};
    \draw[->, black] (\xmin, \m+ 1.5+0.3) -- (\xmax, \m+ 1.5+0.3) node[right] {$t$};

    \draw[->, black] (0,\ymin) -- (0,\ymax+0.3) node[above] {$\Delta$};

    \node at (2.5, \m+0.25) [font=\small] {$I-1$};
    \node at (3, \m+0.75) [font=\small] {$I-2$};

    \foreach \x/\label  in { 0.5/{1}, 1.5/{2}, 2.5/{3}, 3.7/{\ldots}, 5/{$I$}}
        \draw (\x,\m) node[below] [font=\tiny] { \label};

    \foreach \x in {0,1,2,3, 4.5, 5.5}
    \draw (\x,\m+\markheight) -- (\x,\m-\markheight);

    \foreach \y/\label  in { \m+0.5/{1}, \m+1/{2}}
        \draw (-0.2,\y) node[left] [font=\tiny] { \label};

    \foreach \y in {\m+0.5, \m+1}
        \draw (-\markheight, \y) -- (\markheight, \y);

\end{tikzpicture}
     \caption{Illustration to measure the sum VAoI during the inter-delivery time for different
sample path of packet arrival.}
    \label{fig:illustration_to_measure_VAoI}
\end{figure}
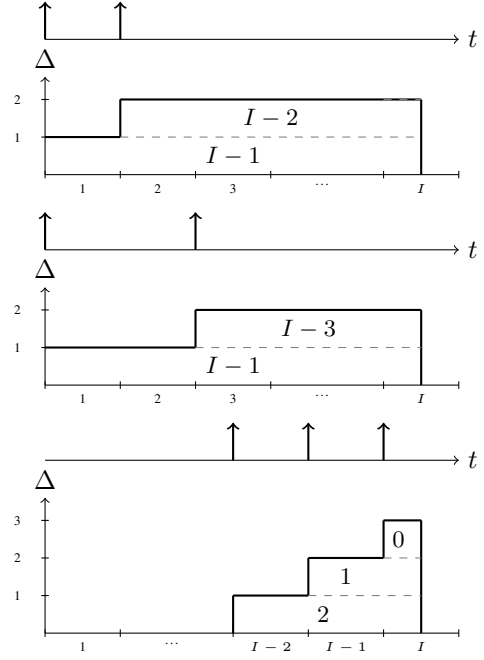
Here, the expectation is  with respect to both the number of packet arrivals $n$, and the positions of these packet arrivals during the inter-delivery time of $I$ slots. Recall that $D_i(T)$ is the total number of deliveries. Given the above expressions, we obtain the lower bound along the lines in \cite{kadota2019_1}. There are $D_i(T)$ inter-delivery slots, denoted as $I_i[1], I_i[2], \ldots, I_i[D_i(T)]$.  Let $R_i$ be the number of remaining slots after the last packet delivery. Here $D_i(T)$, $I_i[1], I_i[2], \ldots, I_i[D_i(T)]$ and $R_i$ are random variables. Then, the time-horizon can be written as follows 
\begin{align}\label{eq:time_horizon}
    T  = \sum_{m = 1}^{D_i(T)} I_i[m] + R_i, \forall i \in \{1, 2,\ldots, N \}.
\end{align}
 The expected sum of VAoI is well-defined within each of the time intervals $I_i[m]$ and $R_i$. According to \eqref{eq:expected_delta}, the expected  sum of VAoI conditioned on $I[m]$ is given by $ \lambda_i(I_i^2[m] - I_i[m] )/2$, for all $m \in \{1, 2,\ldots, D_i(T)\}$, and the expected sum of VAoI conditioned on $R_i$ is $\lambda_i ( R_i^2  - R_i )/2$. Then, the time-average expected sum of VAoI associated with Destination $i$, conditioned on $D_i(T)$, $I_i[1], I_i[2], \ldots, I_i[D_i(T)]$ and $R_i$ can be expressed as
\begin{align}\label{eq:time_average_Vaoi}
         &\frac{1}{T} \sum_{t=1}^{T} \mathbb{E}[\Delta_i(t)]\nonumber\\
         &= \frac{1}{T} \left( \sum_{m = 1}^{D_i(T)} \frac{\lambda_i ( I_i^2[m]  - I_i[m] )}{2} + \frac{ \lambda_i ( R_i^2  - R_i )}{2} \right)\nonumber\\
         & =  \frac{\lambda_i}{2}\left ( \frac{D_i(T)}{T} \frac{1}{D_i(T)} \sum_{m = 1}^{D_i(T)} (I_i^2[m]  - I_i[m]) + \frac{ R_i^2  - R_i}{T} \right).
\end{align}
The sample mean of $ I_i[1], I_i[2],\ldots, I_i[D_i(T)] $ for User $i$ is given by
\begin{align}\label{eq:sample_mean}
    \bar{\mathbb{M}}[I_i] = \frac{1}{D_i(T)} \sum_{m=1}^{D_i(T)} I_i[m].
\end{align}
Employing the sample mean operator $\bar{\mathbb{M}}$ to $I_i^2[m]$ and $I_i[m]$, then applying Jensen's inequality, $\bar{\mathbb{M}}[I_i^2] \geq (\bar{\mathbb{M}}[I_i])^2$, we obtain a lower bound to the time-average expected sum of VAoI as
 \begin{equation}\label{eq:time_average_Vaoi_3}
   \frac{1}{T} \sum_{t = 1}^{T} \mathbb{E}[\Delta_i(t)] \geq \frac{\lambda_i}{2} \left ( \frac{ D_i(T)}{T} \left((\bar{\mathbb{M}}[I_i] )^2 - \bar{\mathbb{M}}[I_i] \right) + \frac{R_i^2 - R_i}{T} \right )
\end{equation}
Then, substituting \eqref{eq:sample_mean} and \eqref{eq:time_horizon} into \eqref{eq:time_average_Vaoi_3}, gives
\begin{align} \label{eq:time_average_Vaoi_4}
    \frac{1}{T} \sum_{t = 1}^{T} \mathbb{E}[\Delta_i(t)] \geq \frac{\lambda_i}{2T} \left( \frac{(T-R_i)^2}{D_i(T)} - (T-R_i) + R_i^2 - R_i \right).
\end{align}
The lower bound, RHS of \eqref{eq:time_average_Vaoi_4}, is independent of set of $I_i[m]$. 
By minimizing the RHS of \eqref{eq:time_average_Vaoi_4} analytically with respect to variable $R_i$, we have $R_i = T/(D_i(T)+1)$. Using this condition, we obtain a lower bound which is independent of $R_i$ as 
\begin{align} \label{eq:time_average_Vaoi_5}
    \frac{1}{T} \sum_{t = 1}^{T} \mathbb{E}[\Delta_i(t)] \geq \frac{\lambda_i}{2} \left( \frac{T}{D_i(T)+ 1} - 1\right).
\end{align}
The lower bound, RHS of \eqref{eq:time_average_Vaoi_5}, is still conditioned on $D_i(T)$, then taking the expectation with respect to $D_i(T)$ and applying Jensen's inequality, gives the updated lower bound as
\begin{align} \label{eq:expected_time_average_vaoi}
    \frac{1}{T} \sum_{t = 1}^{T} \mathbb{E}[\Delta_i(t)] \geq \frac{\lambda_i}{2}\left( \frac{1}{\mathbb{E}\left[\frac{D_i(T)}{T} \right] + \frac{1}{T}} -  1
    \right).
\end{align}
Recall $D_i(T)$ is the total number of packets delivered
to Destination $i$ by the end of the time-horizon $T$ when the policy $\pi \in \Pi$ is employed. The policy $\pi$ gives the rule for obtaining sequence of decisions of transmissions $\{u^{\pi}_i(t)\}^N_{i=1}$ in time slot $t$, then $D_i(T) = \sum_{t=1}^T {u}^{\pi}_i(t)$.
Using the expression of $D_i(T)$ in \eqref{eq:expected_time_average_vaoi}, and  applying the limit $T \to \infty$ gives
\begin{equation}\label{eq:limit_expected_time_average_Vaoi}
    \lim_{T \to \infty} \frac{1}{T} \sum_{t = 1}^{T} \mathbb{E}[\Delta_i(t)] \geq  \frac{\lambda_i}{2}\left( \frac{T}{\mathbb{E}[\sum_{t=1}^T {u}^{\pi}_i(t)]} - 1 \right).
\end{equation}       
The expectation in the above expression is with respect to channel power gain realizations, packet arrival and the policy decisions. Substituting \eqref{eq:limit_expected_time_average_Vaoi} into the objective function in \eqref{eq:main-opt-problem}, yields
\begin{align}\label{eq: expected_vaoi_E[j]_1}
    V_{\rm opt} \geq \frac{1}{2} \sum_{i=1}^{N}  w_i \lambda_i \left(  \frac{T}{\mathbb{E}[\sum_{t=1}^T {u}^{\pi}_i(t)]} - 1 \right).
\end{align}
We define a lower bound $L_B$ as follows: 
\begin{subequations}\label{eq:lower_bound_opt_problem_1}
\begin{align}
    L_B  = \min_{\pi \in \Pi }&\;\frac{1}{2} \sum_{i=1}^{N}  w_i \lambda_i \left(  \frac{T}{\mathbb{E}[\sum_{t=1}^T {u}^{\pi}_i(t)]} - 1 \right),\\
    \text{subject to} &\;\;  \eqref{eq:power_constraint}.
\end{align}
\end{subequations}   
Note that $\mathbb{E}[\sum_{t=1}^T {u}^{\pi}_i(t)]$ is the expected number of successful packet deliveries. In line with Lemma $8$ in \cite{JSAC-Bhat}, we can show  that the lower bound has a stationary randomized policy, similar to the proposed CO-SRP as the optimal policy, and  $\mathbb{E}[\sum_{t=1}^T {u}^{\pi}_i(t)]/T = \hat{q}^{L_B}_i = p_i^{\boldsymbol{\mu}}$. 
By using the definition of $L_B$, $V_{\rm opt}$, and $V_{\rm SRP}$ from \eqref{eq:lower_bound_opt_problem_1}, \eqref{eq:main-opt-problem}, and \eqref{eq:SRP_new}, respectively,  we have,  $L_B  \leq V_{\rm opt} \leq  V_{\rm SRP} $,  i.e.,
\begin{align}\label{eq: a_opt_and_a_srp_relation}
    \frac{1}{2}  \sum_{i = 1}^N 
 w_i \lambda_i \left(\frac{1}{\hat{q}^{L_B}_i} - 1 \right)  \leq V_{\rm opt} \leq \sum_{i = 1}^N w_i \lambda_i \left( \frac{1}{p^{\boldsymbol{\mu}}_{i}} -  
 1 \right).
\end{align}
Noting $\hat{q}^{L_B}_i = p^{\boldsymbol{\mu}}_i$, we conclude that $L_B \leq V_{\rm SRP}/2$. Hence, $V_{\rm SRP} \leq 2 V_{\rm opt}$.

\balance

 \subsection{Proof of Theorem~\ref{thm:MDP}}
 For brevity, we drop $\gamma$ in the notation.
 Suppose it is optimal to transmit $i^{\rm th}$ stream for some $\boldsymbol{\Delta_{-i}}$, $\Delta_i$, $\mathbf{h}$ and $\mathbf{u}_{-i}$. This means that 
\begin{align}
Q(\Delta_i, 1,\cdot) - Q(\Delta_i, 0,\cdot) <0,  
\end{align}
where we replace $\boldsymbol{\Delta_{-i}}$, $\mathbf{h}$, and $\mathbf{u}_{-i}$ with $(\cdot)$ for brevity. Then, it is sufficient to show the following to prove the first part of the theorem: $Q(\Delta_i+1, 1,\cdot) - Q(\Delta_i+1, 0,\cdot) <0$, which prove below. Consider
\begin{align*}
&Q(\Delta_i+1, 1,\cdot) - Q(\Delta_i+1, 0,\cdot)=\nonumber  \\
&  \mathbb{E}_{\mathbf{A}}\left[C(\Delta_i+1, 1,\cdot)\right] + \gamma \mathbb{E}_{\mathbf{A}, \mathbf{H}}\left[V(0, \boldsymbol{\Delta'_{-i}}, \mathbf{h}'|\Delta_i+1,1,\cdot)\right]- \nonumber\\
&   \mathbb{E}_{\mathbf{A}}\left[C(\Delta_i+1, 0,\cdot)\right] - \gamma\mathbb{E}_{\mathbf{A}, \mathbf{H}}\left[V(\Delta_i', \boldsymbol{\Delta'_{-i}}, \mathbf{h}'|\Delta_i+1,0,\cdot)\right]\nonumber\\
& =  0 +\theta P_i+ \sum_{j \neq i}\mathbb{E}_{\mathbf{A}}\left[w_j\Delta'_j+  \theta u_jP_j|\Delta_i+1,1,\cdot\right]\nonumber\\
& + \gamma\mathbb{E}_{\mathbf{A}, \mathbf{H}}\left[V(0, \boldsymbol{\Delta}'_{-i}, \mathbf{h}'|\Delta_i+1,1,\cdot)\right]\nonumber\\ 
& -w_i\lambda_i(\Delta_i+2) -w_i(1-\lambda_i)(\Delta_i+1)\nonumber\\
&-   \sum_{j \neq i}\mathbb{E}_{\mathbf{A}}\left[w_j \Delta''_j+ \theta u_jP_j|\Delta_i+1,0,\cdot\right]\nonumber\\
&-\gamma\mathbb{E}_{\mathbf{A}, \mathbf{H}}\left[V(\Delta_i', \boldsymbol{\Delta'_{-i}}, \mathbf{h}'|\Delta_i+1,0, \cdot)\right]\nonumber\\
&\stackrel{(a)}{\leq}  0 +\theta P_i+ \sum_{j \neq i}\mathbb{E}_{\mathbf{A}}\left[w_j\Delta'_j+ \theta u_jP_j|\Delta_i,1,\cdot\right]\nonumber\\
& \qquad+\gamma\mathbb{E}_{\mathbf{A}, \mathbf{H}}\left[V(0, \boldsymbol{\Delta'_{-i}}, \mathbf{h}'|\Delta_i,1,\cdot)\right] \nonumber\\
& \qquad  -w_i\lambda_i(\Delta_i+1) - w_i(1-\lambda_i)(\Delta_i) \nonumber\\
& \qquad -\sum_{j \neq i}\mathbb{E}_{\mathbf{A}}\left[w_j\Delta'_j+\theta u_jP_j|\Delta_i,0,\cdot\right]\nonumber\\
&\qquad-\gamma\mathbb{E}_{\mathbf{A}, \mathbf{H}}\left[V(\Delta_i', \boldsymbol{\Delta'_{-i}}, \mathbf{h}'|\Delta_i,0, \cdot)\right]\nonumber\\
& = \mathbb{E}_{\mathbf{A}}\left[C(\Delta_i, 1,\cdot)\right] + \gamma \mathbb{E}_{\mathbf{A}, \mathbf{H}}\left[V(0, \boldsymbol{\Delta'_{-i}}, \mathbf{h}'|\Delta_i,1,\cdot)\right] \nonumber\\
&   - \mathbb{E}_{\mathbf{A}}\left[C(\Delta_i, 0,\cdot)\right] - \gamma\mathbb{E}_{\mathbf{A}, \mathbf{H}}\left[V(\Delta_i', \boldsymbol{\Delta'_{-i}}, \mathbf{h}'|\Delta_i,0,\cdot)\right]\nonumber\\
&=Q(\Delta_i, 1,\cdot) - Q(\Delta_i, 0,\cdot), \nonumber
\end{align*}
where $\mathbb{E}_{\mathbf{A}}[\cdot]$ denotes the expectation with respect to the update arrival distribution, $\mathbf{A}$,  and (a) is because of the following: 
\begin{itemize}
    \item[(i)]  When an update from $i^{\rm th}$ stream is transmitted, irrespective of whether the VAoI component of the state is $\Delta_i$ or $\Delta_{i}+1$, the next VAoI corresponding to $i^{\rm th}$ stream will be zero and VAoIs of other streams evolve in the similar manner, and for a given channel power gain vector, $\mathbf{h}$, the decoding order and the transmit powers assigned to each stream remain the same,  and hence
\begin{align*}
&0 +\theta P_i+ \sum_{j \neq i}\mathbb{E}_{\mathbf{A}}\left[w_j\Delta'_j+ \theta u_jP_j|\Delta_i+1,1,\cdot\right] \nonumber\\
& = 0 +\theta P_i+ \sum_{j \neq i}\mathbb{E}_{\mathbf{A}}\left[w_j\Delta'_j+\theta u_jP_j|\Delta_i,1,\cdot\right], 
\end{align*}
and $\mathbb{E}_{\mathbf{A}, \mathbf{H}}\left[V(0, \boldsymbol{\Delta}'_{-i}, \mathbf{h}'|\Delta_i+1,1,\cdot)\right]=\mathbb{E}_{\mathbf{A}, \mathbf{H}}\left[V(0, \boldsymbol{\Delta}'_{-i}, \mathbf{h}'|\Delta_i,1,\cdot)\right]$ hold. 
\item[(ii)] $-w_i\lambda_i(\Delta_i+2) - w_i(1-\lambda_i)(\Delta_i+1)<-w_i\lambda_i(\Delta_i+1) - w_i(1-\lambda_i)\Delta_i$ for any $0\leq \lambda_i\leq 1$ and $\Delta_i\geq 0$. 
\item[(iii)] $\sum_{j \neq i}\mathbb{E}_{\mathbf{A}}\left[w_j\Delta''_j+ \theta u_jP_j|\Delta_i+1,0,\cdot\right]=\sum_{j \neq i}\mathbb{E}_{\mathbf{A}}\left[w_j\Delta''_j+\theta u_jP_j|\Delta_i,0,\cdot\right]$ as  the VAoIs of all streams except $i^{\rm th}$ stream evolve similarly, and transmit powers will be same, irrespective of whether VAoI for $i^{\rm th}$ stream is $\Delta_i$ or $\Delta_i+1$  and 
\item[(iv)] $\mathbb{E}_{\mathbf{A}, \mathbf{H}}[V(\Delta_i', \boldsymbol{\Delta'_{-i}}, \mathbf{h}'|\Delta_i+1, 0, \cdot)]\geq \mathbb{E}_{\mathbf{A}, \mathbf{H}}[V(\Delta_i', \boldsymbol{\Delta'_{-i}}, \mathbf{h}'|\Delta_i, 0, \cdot)]$.
\end{itemize}

We prove (iv) in the following lemma. 
\begin{lemma}
    $\mathbb{E}_{\mathbf{A}, \mathbf{H}}[V(\Delta_i', \boldsymbol{\Delta'_{-i}}, \mathbf{h}'|\Delta_i+1, 0, \cdot)]\geq \mathbb{E}_{\mathbf{A}, \mathbf{H}}[V(\Delta_i', \boldsymbol{\Delta'_{-i}}, \mathbf{h}'|\Delta_i, 0, \cdot)]$ holds. 
\end{lemma}
\begin{proof}
We first prove that the  value function $\mathbb{E}_{\mathbf{A}, \mathbf{H}}[V(\Delta_i, \boldsymbol{\Delta_{-i}}, \mathbf{h})]$ is non-decreasing in $\Delta_i$  for any given $\boldsymbol{\Delta_{-i}}$ and $\mathbf{h}$ via induction on the following value iteration in \eqref{eq:val_iteration}. 
For $m=0$,   $V_{\gamma,0}(\Delta_i,\cdot)$ is trivially non-decreasing in $\Delta_i$. Now, as the induction hypothesis, suppose that $V_{\gamma,m}(\mathbf{s})$ is  non-decreasing in $\Delta_i$ for any $\boldsymbol{\Delta_{-i}}$ and  $\mathbf{h}$. Clearly, $\mathbb{E}[C(\mathbf{s},\mathbf{u})$ is non-decreasing in $\Delta_i$ for any   $\boldsymbol{\Delta_{-i}}$,  $\mathbf{h}$ and  $\mathbf{u}$.  $\mathbb{E}[V_{\gamma,m}(\mathbf{s'})|\mathbf{s},\mathbf{u}]$ is non-decreasing from the induction hypothesis. Hence, as the minimization preserves the non-decreasing property, we have the result.  
From the above,  we know that the value function is non-decreasing in  $\Delta_i$, when $\boldsymbol{\Delta'_{-i}}$ and $\mathbf{h}'$ are fixed. Hence, we have, 
\begin{align}
    & \mathbb{E}_{\mathbf{A}, \mathbf{H}}\left[V(\Delta_i', \boldsymbol{\Delta'_{-i}}, \mathbf{h}'|\Delta_i+1, 0, \cdot)\right] \nonumber \\
    & = \left(1-\lambda_i\right) \mathbb{E}_{\mathbf{A}, \mathbf{H}}\left[V(\Delta_i+1, \boldsymbol{\Delta'_{-i}}, \boldsymbol{h'}\right]   \nonumber\\
    &\quad\;+\lambda_i \mathbb{E}_{\mathbf{A}, \mathbf{H}}\left[V(\Delta_i+2, \boldsymbol{\Delta'_{-i}}, \boldsymbol{h'}\right] \nonumber\\
    & \geq \left(1-\lambda_i\right) \mathbb{E}_{\mathbf{A}, \mathbf{H}}\left[ V(\Delta_i, \boldsymbol{\Delta'_{-i}}, \boldsymbol{h'}) \right]  \nonumber\\
    &\quad \; +\lambda_i \mathbb{E}_{\mathbf{A}, \mathbf{H}}\left[V(\Delta_i+1, \boldsymbol{\Delta'_{-i}}, \boldsymbol{h'}| \Delta_i, 0, \cdot) \right] \nonumber\\
 &   =\mathbb{E}_{\mathbf{A}, \mathbf{H}}\left[V(\Delta_i', \boldsymbol{\Delta'_{-i}}, \mathbf{h}'|\Delta_i, 0, \cdot)\right].
\end{align}
In the above, the inequality follows because (i) since the actions for all the streams except for $i^{\rm th}$ stream are fixed, $\boldsymbol{\Delta_{-i}}$ evolves identically irrespective of whether the VAoI corresponding to $i^{\rm th}$ stream is $\Delta_i$ or $\Delta_{i}+1$ and because (ii) the packet arrivals and the channel power gains transition independently across slots.  
\end{proof}

The proof for the second part of the theorem can be obtained along the similar lines.

\end{document}